\documentclass[11pt,a4paper]{article}
\usepackage{jheppub}

\usepackage{axodraw}
\usepackage[table]{xcolor}
\usepackage{multirow}
\usepackage{longtable}
\usepackage{microtype}
\DisableLigatures{encoding = *, family = * }

\def\eps{\varepsilon}
\def\2{\ \ }
\def\3{\ \ \ }
\def\4{\qquad}
\def\5{\qquad \qquad}
\def\6{\qquad \qquad \qquad}
\def\7{\qquad \qquad \qquad \qquad}
\def\8{\qquad \qquad \qquad \qquad \qquad}
\def\nn{\nonumber}
\def\nl{\nonumber\\}

\def\bc{}

\def\be{\begin{equation}}
\def\ee{\end{equation}}
\def\ba{\begin{eqnarray}}
\def\ea{\end{eqnarray}}
\def\({\left(}
\def\){\right)}
\def\[{\left[}
\def\]{\right]}
\def\.{\ \ \ .}
\def\,{\ \ \ ,}
\def\;{\ \ \ ;}
\newcommand{\df}[2]{{\displaystyle \frac{#1}{#2}}}

\newcolumntype{x}[1]{%
>{\centering
}m{#1}}%

\title{Z-Sum Approach to Loop Integrals using Taylor Expansion}
\author[a]{Paulo A. Rottmann}
\author[a,b]{and Laura Reina}
\affiliation[a]{Department of Physics, Florida State University,\\
513 Keen Building, Tallahassee, Florida, 32306, USA}
\affiliation[b]{KITP, University of California Santa Barbara, CA 93106-4030, USA}
\emailAdd{pr03d@hep.fsu.edu}
\emailAdd{reina@hep.fsu.edu}
\abstract{We study the applicability of the Z-Sum approach to multi-loop calculations with massive particles in perturbative quantum field theory. We systematically analyze the case of one-loop scalar integrals, which represent the building blocks of any higher-loop calculation. We focus in particular on triangle one-loop integrals and identify strengths and limitations of the Z-Sum approach, extending our results to the case of one-loop box integrals when appropriate. We conclude with the calculation of a specific physical example: the calculation of heavy-flavor corrections to deep-inelastic-scattering structure functions.}
\begin{document}
\hfill NSF-KITP-11-122 \vspace{-5mm}

\maketitle

\section{Introduction}

In perturbative quantum field theory it is necessary to deal with loop diagrams when trying to perform higher-order corrections to scattering amplitudes. Regardless of the method being used, results are expressed in a basis of scalar loop integrals which need to be calculated. For single-loop calculations, all integrations are by now known, however such is not the case for multi-loop calculations. While numerical methods can sometimes be useful, ultimately analytical results are desirable. Furthermore, because of the growth in the number of loop integrals to be calculated as more complex diagrams are considered, a fully systematic procedure would be ideal since it  would be more suitable for implementation as an automatic computer package.

With that in mind, the Z-Sum approach has been proposed and is a very strong candidate \cite{Moch:2001zr,Weinzierl:2003jx}. Z-Sums consist of concatenated sums with a particular structure and are useful because they generalize several other important functions including multiple polylogarithms. Furthermore, they form a Hopf algebra including several operations like product, convolution, and conjugation, which are instrumental in simplifying complicated summations.

The method consists of two major steps. First the loop-momentum integration is expressed in terms of concatenated sums. Second, if the concatenated sums match a given pattern, they are systematically reduced to Z-Sums and ultimately multiple polylogarithms, in a procedure that iteratively reduces each individual sum from the innermost all the way to the outermost one.

For multi-loop calculations, there are different options for performing the initial step, using either a Taylor or a Mellin-Barnes expansion. In the last decade, there has been a fair amount of interest in the method, mainly using Mellin-Barnes expansions and applying it successfully to specific calculations \cite{Moch:2002rq,Moch:2002wt,Bierenbaum:2003ud,Bierenbaum:2005:T,Blumlein:2006mh,Bierenbaum:2006mq,Bierenbaum:2007dm,Bierenbaum:2007pn,Bierenbaum:2007qe,Bierenbaum:2008yu,Bierenbaum:2009zt,Bierenbaum:2010jp,Bejdakic:2008cr,Heinrich:2009be,Weinzierl:2009ms,Weinzierl:2009nz,Bolzoni:2009ye,Weinzierl:2010cw,Huber:2010fz,Ablinger:2010ha,DelDuca:2010zp,DelDuca:2010zg}. 

Regardless of the expansion method being used, the full multi-loop integration is performed using unphysical lower-loop integrations as building blocks, for instance three-loop integrations are expressed using two-loop integrations which themselves use one-loop integrations. While in previous works a lot of emphasis has been put on the application of Z-Sum to individual calculations, in this work for the first time we pay particular attention to the procedure of systematically obtaining concatenated sums, necessary for application of Z-Sum algorithms, with a focus on Taylor expansions. We also perform a thorough survey of all physical and unphysical one-loop triangle integrations as building blocks in multi-loop calculations, including massive particles, with the aim of understanding how often the method can be successfully applied, and in which way it needs to be complemented in order to solve a general loop integration. This allows us to identify the key missing algorithms in the Z-Sum machinery, paving the way for an extension of the method. We also for the first time present results for two-loop calculations involving massive internal particles using the Z-Sum reduction algorithms and Taylor series as the expansion method.

The layout of the paper is as follow. We start in section \ref{Loop Integrals and Expansions} with a discussion of the alternative ways in which the Z-Sum algorithms can be applied to multi-loop integrals, depending on the use of either Taylor or Mellin Barnes expansions in intermediate steps. We then focus on Taylor expansions and present a master formula for the method. In section \ref{Gamma Function Expansion and Z-Sums} we cover Z-Sum properties and algorithms, and discuss some missing steps required for a generalization of the method. In section  \ref{Triangle Loop Integrals} we develop a thorough survey of its applicability on one-loop triangle integrals used as building blocks in multi-loop calculations, and in section \ref{Application} we conclude by applying it to specific physically motivated calculations involving two-loop massive integrals necessary for the calculation of heavy-flavor corrections to deep inelastic scattering structure functions. Conclusions are presented in section \ref{Conclusions}. Appendix \ref{Notation and Conventions} lists notations and conventions, while appendix \ref{SumsReordering} presents useful identities for manipulation of concatenated sums.

\section{Loop integrals and expansions}
\label{Loop Integrals and Expansions}
In this section we discuss how to expand loop integrations in order to apply the Z-sums algorithms, and then focus on one particular method based on Taylor expansions.

\subsection{Alternative expansion methods}
A general multi-loop scalar integration can be written as:
\be
\label{TriangleScalar}
\int \frac{d^D k_1}{(2\pi)^D} \ldots \frac{d^D k_r}{(2\pi)^D} \frac{1}{
P_1^{\nu_1} \ldots P_s^{\nu_s}  
}\,
\ee
where the $P_i$ ($i=1,\ldots,s$) represent the denominators of generic loop propagators and have the form $P_i=\left( K_i^2 - m_i^2 \right)$, with $K_i$ a function of the integration variables $k_j$ ($j=1,\ldots,r$) and the external momenta, and $m_i$ are the masses of particles propagating in the loops. Moreover, we assume that both ultraviolet and infrared divergencies in the loop integrations are regularized in dimensional regularization, with $D=2m-2\eps$, where $2m$ is the original space-time dimension and $\eps$ is an infinitesimal.

In order to use the Z-Sum approach, we must express this integral in terms of concatenated sums. This can be achieved by performing a denominator expansion either at the level of the momentum integration, where results of each momentum integration are expressed in terms of propagators in the following momentum integration, or at the level of the parameter integration, after the momentum integration has been performed. There are two alternative methods to perform the expansion, using either a Taylor expansion:
\be
\label{DenominatorTaylor}
\frac{1}{(A+B)^c} = \frac{1}{A^c\ \Gamma(c)} \sum_{n=0}^\infty \frac{\Gamma(n+c)}{\Gamma(n+1)} \(\frac{-B}{A}\)^n\,
\ee
or the inverse Mellin-Barnes transformation \cite{Smirnov:2004,Czakon:2005rk}:
\be
\label{DenominatorMB}
\frac{1}{(A+B)^c} = \frac{1}{2 \pi i} \frac{1}{A^c\ \Gamma(c)} \int_{-i \infty}^{i \infty} \Gamma(z+c) \Gamma(-z) \(\frac{B}{A}\)^z\.
\ee
We refer to \ref{DenominatorMB} as an expansion, even though it is not, because it will lead to a series after the complex integration is performed by completing the contour over complex infinity and applying the residue theorem.

As a result of the application of these expansions, the loop integral in eq. (\ref{TriangleScalar}) is expressed in terms of concatenated sums through a series of simple steps, initially expressing the momentum integration as a parameter integral and finally reducing the latter to beta functions, thus systematizing the calculation of both the momentum and parameter integrations. While the Taylor expansion may only be applied to the parameter integration, the Mellin-Barnes transformation can be used either at the momentum or the parameter-integration level, leading to three interesting distinct approaches to perform the expansion of multi-loop scalar integrals, which we summarize in table \ref{TableABC}.

\begin{table}[t]
\begin{center}
\begin{tabular}{ x{.4cm} >{\columncolor[rgb]{0.95,0.95,.95}}p{4cm} p{4cm} >{\columncolor[rgb]{0.95,0.95,0.95}}p{4cm} }
&\rule[8.8cm]{0cm}{0cm}&&\\
\end{tabular}
\end{center}
\end{table}
\addtocounter{table}{0}

\begin{table}[t]
\vspace{-9.8 cm}
\begin{center}
\begin{tabular}{x{.4cm} p{4cm} p{4cm} p{4cm} }
\cline{2-4}
&\multicolumn{1}{ x{4cm} }{Approach A}		& \multicolumn{1}{ x{4cm} }{Approach B}	& \multicolumn{1}{ x{4cm} }{Approach C}	\\ \cline{2-4}
1&\multicolumn{3}{ x{12cm} }{Introduce Feynman parameters in first loop} \\ \cline{2-4}
2&\multicolumn{3}{ x{12cm} }{Perform momentum integration of first loop} \\ \cline{2-4}
3&\multicolumn{2}{ x{8cm} }{Express result of initial steps as one ``artificial'' propagator in second loop} & \multicolumn{1}{ x{4cm} }{Using Mellin-Barnes, express result as several propagators in second loop}  \\ \cline{2-4}
4&\multicolumn{3}{ x{12cm} }{Introduce Feynman parameters in second loop} \\ \cline{2-4}
5&\multicolumn{3}{ x{12cm} }{Perform momentum integration of second loop} \\ \cline{2-4}
6&\multicolumn{3}{ x{12cm} }{Repeat procedure for more loops if necessary and perform all momentum integrations} \\ \cline{2-4}
7&\multicolumn{1}{ x{4cm} }{Expand denominator \break using Taylor expansion}		& \multicolumn{1}{ x{4cm} }{Expand denominator using Mellin-Barnes expansion}			& \multicolumn{1}{ x{4cm} }{No expansion necessary}	\\ \cline{2-4}
8&\multicolumn{3}{ x{12cm} }{Perform all parameter integrations} \\ \cline{2-4}
9&\multicolumn{1}{ x{4cm} }{No complex integration} & \multicolumn{2}{x{8cm} }{Perform all complex integrations} \\ \cline{2-4}
10&\multicolumn{3}{ x{12cm} }{Obtain final result in terms of concatenated sums} \\ \cline{2-4}
\end{tabular}
\caption{Steps for multi-loop calculations using approaches A, B and C.\label{TableABC}}
\end{center}
\end{table}

Approaches A and B follow the standard procedure for multi-loop momentum integration, and the expansion is only performed after all momentum integrations are done, 
where the factor being expanded is the denominator of the parameter integration. In order for the expansion to be legal, conditions must be met. For approach A, which uses Taylor expansions, $(A/B)$ (see eq. (\ref{DenominatorTaylor})) will involve physical invariants and integration parameters, and the condition $|(A/B)|<1$ must be enforced for some choice of invariants (physical or not) over the whole range of integration. Conditions for approach B are only enforced when the complex integration is performed by completing the integration contour and using the residue theorem, with the result being a (concatenated) sum over residues. In order to complete the contour, the integrand must vanish at complex infinity, which will lead to the condition of convergence. 

Approaches A and B share many of the same steps and thus are very similar. All well defined expansions using approach A are equivalent to the cases in approach B where the result from the parameter integrations does not contain poles. Approach B is more general however, since it also allows for cases where the result from the parameter integration involves poles, unlike approach A, where these cases would correspond to badly defined expansions.

Approach C is different from the previous two as it introduces Mellin-Barnes transformations before all momentum integrations have been performed, with the expansion being used to express the results of one momentum integration in terms of propagators of the next momentum integration, and not at the level of the parameter integrations. Unlike the previous case, Taylor expansions cannot be used in place of Mellin-Barnes transformations as it would always lead to badly defined expansions.

While these methods have been successfully applied to specific calculations
\cite{Moch:2002rq,Moch:2002wt,Bierenbaum:2003ud,Bierenbaum:2005:T,Blumlein:2006mh,Bierenbaum:2006mq,Bierenbaum:2007dm,Bierenbaum:2007pn,Bierenbaum:2007qe,Bierenbaum:2008yu,Bierenbaum:2009zt,Bierenbaum:2010jp,Bejdakic:2008cr,Heinrich:2009be,Weinzierl:2009ms,Weinzierl:2009nz,Bolzoni:2009ye,Weinzierl:2010cw,Huber:2010fz,Ablinger:2010ha,DelDuca:2010zp,DelDuca:2010zg}, a thorough study of each approach has not yet been performed. Such undertake would be necessary in order to understand the types of summations obtained from a general loop integration and to learn how often the Z-Sum approach is applicable. While we would like to eventually study all three methods in order to find out which one is most promising, in this work we will focus on approach A.

\subsection{General expression for approach A}
We will now derive a general expression for approach A. Following the first six steps as shown in table \ref{TableABC}, Feynman parameters are introduced and all momentum integrations are performed. We obtain a general parameter integration of the form:
\be
\label{GeneralConcatenatedSumInitial}
I=\int_0^1 \frac{
\prod_{l=1}^{n} du_l\ u_l^{a_{(2l-1)}} (1-u_l)^{a_{(2l)}}
}{
\prod_{k=1}^\gamma \left(1-\sum_{t_k=1}^{\beta_k} c_{(k,t_k)} \prod_{s=1}^n u_s^{p_{(k,t_k,2s-1)}} (1-u_s)^{p_{(k,t_k,2s)}}\right)^d
}\ ,
\ee
where we have performed a change of variable so that
all parameter integration variables $u_l$ ($l=1,\ldots,n$) range from $0$ to $1$, and we have omitted any overall factor not relevant to the Z-Sum procedure, even if it involves an infinitesimal part, since this is not crucial to the development of our discussion. In eq. (\ref{GeneralConcatenatedSumInitial}), $n$ is the number of parameter variables (equal to the number of propagators minus 1), $\gamma$ is the number of factors in the denominator, and the coefficients $c_{(k,t_k)}$ involve invariants (dot products between external momenta and masses squared). The powers in the denominator terms must obey $p_{(k,t_k,s)} \ge 0$. The form $(1-\Delta)$ in the denominator can always be achieved by simple manipulations if not immediately obtained from the previous steps. Note that the form of the polynomial in the denominator is not unique since we are expressing it in powers of both $u_s$ and $(1-u_s)$ ($s=1,\ldots,n$), with different forms leading to different (but equivalent) summations when expanded.

We proceed with the Taylor expansion of the denominator followed by binomial expansions if necessary, after which all parameter integrations are of the form of beta functions. All these steps can be summarized in the following master formula:
\ba
\label{GeneralConcatenatedSumAfterInt}
I&=& \frac{1}{\Gamma(d)^\gamma} 
\prod_{k=1}^\gamma \sum_{i_{(k,0)}=0}^{\infty} \frac{\Gamma(i_{(k,0)}+d)}{\Gamma(i_{(k,0)}+1)}
 \sum_{i_{(k,1)}=0}^{i_{(k,0)}} \ldots \sum_{i_{(k,\beta_k-1)}=0}^{i_{(k,\beta_k-2)}}\\ 
&\times& \left(i_{(k,0)}-i_{(k,1)},\ \ldots\ ,i_{(k,\beta_k-1)}-i_{(k,\beta_k)}\right)! 
\prod_{t_k=1}^{\beta_k} c_{(k,t_k)}^{i_{(k,t_k-1)}-i_{(k,t_k)}}
\( \prod_{l=1}^n \frac{\Gamma(g_{(2l-1)}) \Gamma(g_{(2l)})}{\Gamma(g_{(2l-1)}+g_{(2l)})} \)^{\frac{1}{\gamma}}\ ,\nn
\ea
with 
\be
\label{GammaArgs}
g_{(\ell)} = a_{(\ell)}+1+\sum_{q=1}^\gamma \sum_{j_q=0}^{\beta_q-1} i_{(q,j_q)} \(p_{(q,j_q+1,\ell)}-p_{(q,j_q,\ell)}\)\,
\ee
where $p_{(q,0,\ell)}=i_{(k,\beta_k)}=0$ was used to shorten the notation, and we used the multinomial coefficient notation:
\be
(a_1,\ \ldots\ ,a_n)!\ =\ \frac{\Gamma(a_1+\ \ldots\ +a_n+1)}{\Gamma(a_1+1)\ \ldots\ \Gamma(a_n+1)}\ .
\ee

Although eqs. (\ref{GeneralConcatenatedSumAfterInt}) and (\ref{GammaArgs}) may not look very transparent because of their generality, they are extremely useful as they show the types of sums to be expected when using approach A. The most important part of these expressions, and what defines whether the Z-Sum reduction will be possible, is the argument of the gamma functions arising from the parameter integrations, namely eq. (\ref{GammaArgs}). It consists of a constant and a variable part. The latter depends on the summation variables $i_{(q,j_q)}$ multiplied by the exponents $p_{(q,j_q,\ell)}$ of the terms in the original polynomial in the denominator of the parameter integration. Depending on the form of this denominator, the arguments of the gamma functions will involve several summation variables, and will therefore be more complicated to deal with. The constant term $a_{(\ell)}$ involves integers and possibly an infinitesimal $\eps$ stemming from the $D=2m-2\eps$ of dimensional regularization. 

\section{Gamma function expansion and Z-Sums}
\label{Gamma Function Expansion and Z-Sums}
Whenever an infinitesimal is present in the argument of a gamma function, we use the following expansions:
\ba
\label{GammaExpansion}
\Gamma(n+\epsilon) &=& \theta(n>0)\ \Gamma(1+\epsilon)\ \Gamma(n)\ \sum_{i=0}^{n-1}\ \epsilon^i\ Z_i(n-1)\nl
&+& \theta(n \le 0)\ \frac{\Gamma(1+\epsilon)\ (-1)^n}{\epsilon\ \Gamma(1-n)}\ \sum_{i=0}^{\infty}\ \epsilon^i\ S_i(-n)\,\\
\label{IGammaExpansion}
\frac{1}{\Gamma(n+\epsilon)}&=&\theta(n>0)\ \frac{1}{\Gamma(1+\epsilon)\ \Gamma(n)} \sum_{i=0}^{\infty}\ (-\epsilon)^i\ S_i(n-1)\nl
&+& \theta(n \le 0)\ \frac{\epsilon\ \Gamma(1-n)\ (-1)^n}{\Gamma(1+\epsilon)} \sum_{i=0}^{-n}\ (-\epsilon)^i\ Z_i(-n)\,
\ea
where $n$ is an integer, $\theta$ is a boolean step function, and $Z_i$ and $S_i$ are called Euler-Zagier and harmonic sums, respectively. These functions are special cases of Z-Sums and S-Sums \cite{Moch:2001zr,Weinzierl:2003jx,Weinzierl:2004bn,Rottmann:2011Diss}, defined by:
\begin{align}
Z(n) &=\theta(n \ge 0),\qquad
&   
Z(n;m_1,\ldots,m_k;x_1,\ldots,x_k) =
\sum_{n\ge i_1>i_2>\ldots>i_k>0} \frac{x_1^{i_1}}{i_1^{m_1}} \ldots \frac{x_k^{i_k}}{i_k^{m_k}}  \ ,\\   
S(n) &= \theta(n > 0),\qquad
&
S(n;m_1,\ldots,m_k;x_1,\ldots,x_k) =
\sum_{n\ge i_1\ge i_2\ge \ldots\ge i_k\ge1} \frac{x_1^{i_1}}{i_1^{m_1}} \ldots \frac{x_k^{i_k}}{i_k^{m_k}}\ ,
\end{align}
where $k$ is the ``depth'' and $(m_1+\ldots+m_k)$ is the ``weight'' of a sum. Z-Sums also generalize other functions including multiple polylogarithms of Goncharov, harmonic polylogarithms of Remiddi-Vermaseren, Nielsen's generalized polylogarithms, and classical polylogarithms \cite{Euler:1775:M,Goncharov:1998:M,Lewin:1981:P,Nielsen:1909:D,Remiddi:1999ew}. Z-Sums satisfy a Hopf algebra, some properties of which are illustrated in table \ref{Z-Sums Properties}. More details on these operations can be found in \cite{Moch:2001zr,Rottmann:2011Diss}.

Note that, at first look, the building blocks of the algorithms in table \ref{Z-Sums Properties} match the ones obtained in eq. (\ref{GeneralConcatenatedSumAfterInt}), namely binomial coefficients, $x^i$ factors, inner Z-Sums, and $\frac{1}{a_1i+b_1}$ factors originating from the simplification of ratios of gamma functions. The intent of the method is to use the properties of Z-Sums to systematically reduce the expression obtained from the expansion of loop integrals to multiple polylogarithms.

{\renewcommand{\arraystretch}{1.5}
\begin{table}[tp]
\begin{center}
\begin{tabular}{l@{ }r@{ }l}
\hline 
Conversion			& $Z(n,\ldots)$		& $\leftrightarrow S(n,\ldots)$\\ \hline 
Limit shifting			& $Z(n\!+\!c,\ldots)$		& $\leftrightarrow Z(n,\ldots)$\\ \hline 
Limit multiplication		& $Z(2n,\ldots)$		& $\leftrightarrow Z(n,\ldots)$\\ \hline 
Multiplication			& $Z(n,\ldots)Z(n,\ldots)$		& $\leftrightarrow Z(n,\ldots)$\\ \hline 
Conjugation			& $\displaystyle{\sum_{i=1}^n \binom{n}{i} \frac{x_1^i}{\left(i\!+\!b\right)^{m}} S(i\!+\!d_1,\ldots) }$		& $\leftrightarrow S(n,\ldots)$\\ \hline 
\multicolumn{3}{ l }{Convolution} \\
\multicolumn{2}{ r }{$\displaystyle{\sum_{i=1}^n \frac{x_1^i}{\left(a_1 i\!+\!b_1\right)^{m_1}} Z(i\!+\!d_1,\ldots) \frac{x_2^i}{\left(a_2 \left(n\!-\!i\right)\!+\!b_2\right)^{m_2}}  Z(n\!-\!i\!+\!d_2,\ldots)}$}		& $\leftrightarrow Z(n,\ldots)$\\ \hline 
\multicolumn{3}{ l }{Convolution and conjugation}\\
\multicolumn{2}{ r }{$\displaystyle{\sum_{i=1}^n \binom{n}{i} \frac{x_1^i}{\left(i\!+\!b_1\right)^{m_1}} Z(i\!+\!d_1,\ldots) \frac{x_2^i}{\left(n\!-\!i\!+\!b_2\right)^{m_2}}  Z(n\!-\!i\!+\!d_2,\ldots)}$} & $\leftrightarrow Z(n,\ldots)$ \\ \hline 
\end{tabular}
\caption{Properties of Z and S-Sums.\label{Z-Sums Properties}}
\end{center}
\end{table}
}

\subsection{General simplifications}
\label{General Simplifications}
Before one can use the Z-Sum algorithms listed in table \ref{Z-Sums Properties} on special cases of eq. (\ref{GeneralConcatenatedSumAfterInt}), some simplifications are necessary. Some of these are trivial but are included for completeness.

\paragraph{Ratios of gamma functions} After the expansion in powers of $\eps$ has been performed, one still needs to simplify ratios of gamma functions in order to use the Z-Sum algorithms. This can be done using:
\be
\label{PartialFractioningofGammaPair}
\frac{\Gamma(i+a)}{\Gamma(i+b)} = \frac{1}{(i+b-1)(i+b-2)\ldots(i+a)} =
\frac{1}{\Gamma(b-a)} \sum_{j=0}^{b-a-1} \binom{b-a-1}{j} \frac{(-1)^j}{i+a+j}\ ,
\ee
if $(b-a)$ is a positive integer and:
\be
\label{PartialFractioningofGammaPair2}
\frac{\Gamma(i+a)}{\Gamma(i+b)} = (i+a-1)(i+a-2) \ldots (i+b) = \sum_{j=0}^{a-b} \kappa_{a-b-j}(b,\ldots,a-1)\ i^j\ ,
\ee
if $(b-a)$ is a negative integer, where we define the kappa function $\kappa_s(t)$ as the sum of products of every subset of $t$ with $s$ elements, for example:
\ba
\kappa_0(\ldots)&=&1\, \nl
\kappa_1({a,b,c})&=&a+b+c\, \nl
\kappa_2({a,b,c})&=&\(a b+ a c+ b c\)\, \\
\kappa_3({a,b,c})&=&abc\. \nn
\ea

\paragraph{Partial Fractioning} If more than one pair of gamma functions exist, one needs to perform partial fraction using:
\ba
\label{GeneralPartialFractioning}
&&\4 \frac{1}{(a i+b)^{m_1}}\ \frac{1}{(c i+d)^{m_2}}=\nl
&&=\(\frac{a}{a d-c b}\)^{m_2} \sum_{j=1}^{m_1} \binom{m_1+m_2-j-1}{m_2-1} \(\frac{-c}{a d-c b}\)^{m_1-j} \frac{1}{(a i+b)^j}\\
&&+\(\frac{-c}{a d-c b}\)^{m_1} \sum_{j=1}^{m_2} \binom{m_1+m_2-j-1}{m_1-1} \(\frac{a}{a d-c b}\)^{m_2-j} \frac{1}{(c i+d)^j}\,\nn
\ea
where $m_1$, $m_2$, $a$, and $c$ are integers.

\paragraph{Lowering and Raising Operators}In some cases we need to perform sums involving powers of the summation variable in the numerator. Sometimes it is useful to remove these factors by using derivatives, as in: 
\be
\sum_{i=1}^n i^m x^i f(i) = \(x \frac{d}{dx}\)^m \left( \sum_{i=1}^n x^i f(i) \right)\ .
\ee
Similarly, when the $i^m$ factor is in the denominator we could use:
\be
\sum_{i=1}^n \frac{x^i f(i)}{i^m} = \underbrace{\int \frac{dx'}{x'} \ldots \int \frac{dx'}{x'}}_m \( \sum_{i=1}^n (x')^i f(i) \)\ .
\ee
In case $f(i)$ also depends on $x$ we can always make the substitution $x^i \to (x_0 x)^i$ (but not in $f(i)$), perform the derivative or integration with respect to $x_0$ and ultimately take the limit $x_0 \to 1$. These are called ``lowering'' and ``raising'' operators \cite{Weinzierl:2003jx}: 
\ba
\label{RaiseLowerOperators}
(x^-)\cdot f(x) & = & x \frac{d}{dx} f(x)\, \nl
(x^+)\cdot f(x) & = & \int_0^x \frac{dx'}{x'} f(x')\, \\
(x^+)^m\cdot 1 & = & \frac{1}{m!} \ln^m(x)\. \nn
\ea

\paragraph{Reversing summation order}
If a sum involves the summation variable $i$ solely as $(n-i)$, where $n$ is the upper limit, it can be trivially simplified as:
\be
\sum_{i=0}^n f(n-i) = \sum_{i=0}^n f(i)\.
\ee

\paragraph{Shifted sums}
The application of step functions $\theta$ on sums will modify the summation limits. In order to use the Z-Sum algorithms, these need to be shifted back to the standard values. Some examples of such an operation are:
\ba
\sum_{i=0}^n \theta(i \le a) f(i) &=& \theta(n \le a) \sum_{i=0}^n f(i) + \theta(0 \le a < n) \sum_{i=0}^a f(i)\,\nl
\sum_{i=0}^n \theta(i \ge n-a) f(i) &=& \theta(n \le a) \sum_{i=0}^n f(i) + \theta(n > a) \sum_{i=0}^a f(i+n-a)\,\\
\sum_{i=0}^n \theta(i \ge a) f(i) &=& \theta(a \le 0) \sum_{i=0}^n f(i) + \theta(0 < a \le n) \sum_{i=0}^{n-a} f(i+a)\.\nn
\ea

\paragraph{Binomial synchronization}
After sum shifting, the binomial coefficient, if present, might not match the summation's limit. It can be ``synchronized'' using: 
\be
\sum_{i=0}^{n+a} \binom{n}{i+b} f(i+b)=\sum_{i=0}^{n+a} \binom{n+a}{i} \frac{\Gamma(n+1)}{\Gamma(n+a+1)} \frac{\Gamma(i+1)}{\Gamma(i+b+1)} \frac{\Gamma(n-i+a+1)}{\Gamma(n-i-b+1)}  f(i+a)\.
\ee

\subsection{Reduction to Z-Sums algorithms}
Combining the Z-Sum algorithms of table \ref{Z-Sums Properties} with the simplifications of subsection \ref{General Simplifications}, we obtain reduction algorithms for the following summation types:
\begin{align}
\label{AlgA}
\text{A} & \4 \displaystyle{\sum_{i=1}^n \frac{x_1^i}{(a_1 i\!+\!b_1)^{m_1}} \ldots \frac{x_r^i}{(a_r i\!+\!b_r)^{m_r}}\ X(c_1 i\!+\!d_1;\ldots) \ldots X(c_s i\!+\!d_s;\ldots)}\,\\
\label{AlgB}
\text{B} & \4 \displaystyle{\sum_{i=1}^n \frac{x_1^i}{(a_1 f_i^n\!+\!b_1)^{m_1}} \ldots \frac{x_r^{f_i^n}}{(a_r f_i^n\!+\!b_r)^{m_r}}\ X(c_1 f_i^n\!+\!d_1;\ldots) \ldots X(c_s f_i^n\!+\!d_s;\ldots)}\,\\
\label{AlgC}
\text{C} & \4 \displaystyle{\sum_{i=1}^n \binom{n}{i} \frac{x_1^i}{(i\!+\!b_1)^{m_1}} \ldots \frac{x_r^i}{(i\!+\!b_r)^{m_r}}\ X(c_1 i\!+\!d_1;\ldots) \ldots X(c_s i\!+\!d_s;\ldots)}\,\\
\label{AlgD}
\text{D} & \4 \displaystyle{\sum_{i=1}^n \binom{n}{i} \frac{x_1^{f_i^n}}{({f_i^n}\!+\!b_1)^{m_1}} \ldots \frac{x_r^{f_i^n}}{({f_i^n}\!+\!b_r)^{m_r}}\ X(c_1 {f_i^n}\!+\!d_1;\ldots) \ldots X(c_s {f_i^n}\!+\!d_s;\ldots)}\,
\end{align}
where each $X$ represents either a Z-Sum or S-Sum, $b_\ell$ and $d_\ell$ are \emph{bounded} numbers,\footnote{For the purpose of this paper, we will use the term \emph{bounded} numbers to define numbers  that involve integers and variables from summations with a finite number of terms, while we will denote as \emph{unbounded} numbers that involve summation variables going to infinity.} and $f_i^n$ equals either $i$ or $(n-i)$, with at least one of each present for algorithms B and D. These algorithms are similar to those presented in \cite{Weinzierl:2003jx} with the addition of $a_\ell$ and $c_\ell$ multipliers,\footnote{The algorithms in eqs.~(\ref{AlgA})-(\ref{AlgD}), without $a_l$ and $c_l$ multipliers, have been implemented in a very useful FORM~\cite{vanOldenborgh:1990wj,Vermaseren:2000nd} package presented in~\cite{Moch:2005uc}.} and are used iteratively for a systematic reduction of sums obtained from loop integrals, applying them from the innermost sum all the way to the outermost one. We will refer to expressions where algorithms A, B, C, and D are sufficient as having a \emph{seamless reduction}.

\paragraph{Missing Algorithms}
While algorithms A, B, C, and D allow for the solution of many loop integrals, a study of eq. (\ref{GeneralConcatenatedSumAfterInt}) shows that they are not sufficient for the general case when using Taylor expansion. The problem stems from two main limitations, namely a lack of a multiplier $a_\ell$ in algorithms C and D for some parameter integrations with quadratic denominator, and the requirement that the offsets $b_\ell$ and $d_\ell$ be bounded numbers.

If we focus on one limitation at a time, it is possible to list the missing algorithms as given by:
\begin{align}
\label{MA1}\text{\parbox{4cm}{\center Higher Power\break Variations of Type C}} & \4 \displaystyle{\sum_{i=0}^{n_1} {\color{red}\bf \binom{n_1}{i}} \frac{x^i}{({\color{red}\bf a} \ \!i+b)^m} Z(2i,\ldots)}\,\\
\label{MA2}\text{\parbox{4cm}{\center Unbounded Shifting of Z-Sum Upper Limit}} & \4 \displaystyle{\sum_{i=1}^{n_1} \frac{x^i}{i+b} Z(i-1,\ldots) Z(i+{\color{red}\bf n_2},\ldots)}\,\\
\label{MA3}\text{\parbox{4cm}{\center Unbounded Offset \break of Type A}} & \4 \displaystyle{\sum_{i=1}^{n_1} \frac{x^i}{(i+{\color{red}\bf n_2})^m} Z(i-1;\ldots)}\,\\
\label{MA4}\text{\parbox{4cm}{\center Unbounded Offset \break of Type C}} & \4 \displaystyle{\sum_{i=1}^{n_1} \binom{n_1}{i} \frac{x^i}{(i+{\color{red}\bf n_2})^m} S(i;\ldots)}\,
\end{align}
where $b$ is a \emph{bounded} number while $n_1$ and $n_2$ are \emph{unbounded}, and we highlighted in red bold face the stumbling block in each case.

In eq. (\ref{MA1}), we have used a general factor $a$ in the denominator $(a i+b)^m$. This factor comes straight from the powers of $u_s$ and $(1-u_s)$ in the denominator of the parameter integration (see eq. (\ref{GeneralConcatenatedSumInitial})). For the cases involving single-loop triangle integrations, to be discussed in section \ref{Triangle Loop Integrals}, only solutions with $a=2$ are necessary, since those polynomials are at most quadratic.

It is important to note that more missing algorithms exist when combining more than one stumbling block. While some individual calculations involving sums like the ones above can be expressed in terms of multiple polylogarithms when all concatenated sums are taken into account,\footnote{Appendix \ref{SumsReordering} lists several operations useful when performing operations at the summation level.} a systematic and iterative procedure to reduce these types of sums to Z-Sums would be very useful but has not yet been developed, and in fact it is unknown whether it is possible.

\section{Triangle loop integrals}
\label{Triangle Loop Integrals}
In order to study the applicability of approach A we will consider the case of all one-loop triangle integrals. They are the simplest yet non-trivial class of loop integrals and are important since they are building blocks of multi-loop calculations. In order to obtain the results presented in this section we have developed extensive computer algorithms using Mathematica as a programming language. While the results discussed refer specifically to triangle integrations, the same procedure is applicable to more complicated integrations like box integrations and beyond.

\subsection{Reducible loop-like integrals}
We have used two different methods for evaluating triangle loop integrals, which we call ``forward'' and ``backward'' approaches.

\paragraph{Forward approach}When faced with a specific triangle loop integration, one possible way to find out whether the Z-Sum method would be sufficient is to use what we call forward method. In this approach, we consider every possible parametrization of the integration, study different ways of performing the expansion, and verify whether one of the resulting sums matches the Z-Sum algorithms, thus leading to an analytic expression in terms of multiple polylogarithms. In general, each diagram will lead to one or more well defined expansions, some of which may immediately match the structure of the Z-Sums algorithms, thus leading to a \emph{seamless reduction}. For other cases, however, none of the expansions obtained can be immediately reduced. If the result is known by some other integration method, it is always possible to manipulate the expression at the summation level, using the algorithms presented in Appendix \ref{SumsReordering}, to obtain the known result. The steps for performing such manipulation, however, are always different for each case, and a systematic generalization of the procedure, necessary for obtaining new solutions, is still lacking.

The forward method is interesting because it gives the form of summations that cannot be seamlessly reduced, and has been key in understanding what types of necessary algorithms are still missing (eqs. (\ref{MA1}-\ref{MA4})), however it is not very convenient for finding all possible summations that can be seamlessly reduced since it requires a brute force approach.

\paragraph{Backward approach}Another more systematic approach for finding all possible solutions is to consider the Z-Sum algorithms A, B, C, and D as building blocks, and use them to construct all possible concatenated sums that can be seamlessly reduced. We then perform the calculation in reverse order, obtaining in the end all loop-like integrations with systematic reduction. We start this procedure by listing all building blocks, as shown in table \ref{BuildingBlockSums}.

{\renewcommand{\arraystretch}{1.4}
\begin{table}[tp]
\begin{center}
\begin{tabular}{|l|c|}
\hline 
\multicolumn{1}{|c|}{Sum}						&\multicolumn{1}{c|}{Beta Functions}							\\ \hline
$\sum_{i=0}^\infty\ \frac{\Gamma(i+d)}{\Gamma(i+1)\Gamma(d)} x^i$	& $[0,0],\ [i,0],\ [2i,0]$								\\ \hline

$\sum_{i=0}^{n}\ \binom{n}{i}\ x^i$ 					& $[0,0],\ [i,0],\ [n-i,0],\ [i,n-i]$							\\ \hline
$\sum_{i=0}^n\ f(n,i)\ x^i$ 						& $[0,0],\ [i,0],\ [2i,0],\ [n-i,0],\ [2n-2i,0]$					\\ \hline

$\sum_{i=0}^{q-n}\ \binom{q-n}{i}\ x^i$					& $[0,0],\ [i,0],\ [q-n-i,0],\ [i,q-n-i]$						\\ \hline
$\sum_{i=0}^{q-n}\ f(q-n,i)\ x^i$					& $[0,0],\ [i,0],\ [2i,0],\ [q-n-i,0],\ [2q-2n-2i,0]$					\\ \hline

$\sum_{i=0}^{2n}\ \binom{2n}{i}\ x^i$ 					& $[0,0],\ [i,0],\ [2n-i,0],\ [i,2n-i]$							\\ \hline
$\sum_{i=0}^{2n}\ f(2n,i)\ x^i$						& $[0,0],\ [i,0],\ [2i,0],\ [2n-i,0],\ [4n-2i,0]$					\\ \hline

$\sum_{i=0}^{2q-2n}\ \binom{2q-2n}{i}\ x^i$	 			& $[0,0],\ [i,0],\ [2q-2n-i,0],\ [i,2q-2n-i]$						\\ \hline
$\sum_{i=0}^{2q-2n}\ f(2q-2n,i)\ x^i$					& $[0,0],\ [i,0],\ [2i,0],\ [2q-2n-i,0],\ [4q-4n-2i,0]$					\\ \hline

\end{tabular}
\caption{Single sums used as building blocks, obtained from Z-Sum algorithms A, B, C, and D (eqs. (\ref{AlgA}-\ref{AlgD})). Each bracket $[p_1,p_2]$ represents a beta function (see eq.~(\ref{betafunction})), and so the correct number
of bracket needs to be selected corresponding to the number of parameter integrations (2 for triangles, 3 for boxes, etc). See eqs. (\ref{x1Explanation}) and (\ref{xExplanation}) for the relation between $x$ and $[p_1,p_2]$. In the last two rows $q$ represents the upper limit in the hidden $n$ sum. $f(j,i)$ stands for any of the following: $1$, $\frac{\Gamma(i+d)}{\Gamma(i+1)\Gamma(d)}$, $\frac{\Gamma(j-i+d)}{\Gamma(j-i+1)\Gamma(d)}$ or $\frac{\Gamma(j-i+d)}{\Gamma(j-i+1)\Gamma(d)} \frac{\Gamma(i+d)}{\Gamma(i+1)\Gamma(d)}$, where $d$ will become the denominator power in the loop-like integration. \label{BuildingBlockSums}}
\end{center}
\end{table}
}

Each variable $x^i$ in table \ref{BuildingBlockSums} has one of the following forms:
\be
\label{x1Explanation}
c^i,\4 c^i [p_1,p_2],\4 c^i [p_1,p_2] [p_3,p_4],\4 c^i [p_1,p_2] [p_3,p_4] \ldots\,
\ee
where $c^i$ is a coefficient, while $[p_j,p_k]$ can be chosen from the corresponding line in the second column of table \ref{BuildingBlockSums} and represent a beta function defined by:
\be
\label{betafunction}
[p_j,p_k]=\frac{\Gamma(p_j+a_j)\Gamma(p_k+a_k)}{\Gamma(p_j+p_k+a_{jk})}= \int_0^1\ du\ u^{p_j+a_j-1}\ (1\!-\!u)^{p_k+a_k-1}\,
\ee
with $a_j$ and $a_k$ constants. If we consider the product of all $x$ variables chosen, we obtain:
\be
\label{xExplanation}
x_1^{i_1} \ldots x_s^{i_s}=c_1^{i_1} \ldots c_s^{i_s} \times [p_1,p_2] \ldots [p_{2b-1},p_{2b}]\ ,
\ee
where the subscripts in $x^i$ and $c^i$ differentiate between the individual sums picked as building blocks, $s$ is the total number of sums, $b$ is the desired number of parameter integrations, and the product $[p_1,p_2] \ldots [p_{2b-1},p_{2b}]$ consists of a combination of any of the brackets allowed in each building block.

While there are other building-block sums that lead to a \emph{seamless reduction}, for example:
\be
\sum_{i=0}^{\lfloor \frac{n}{2} \rfloor}\ \binom{\lfloor \frac{n}{2} \rfloor}{i}\ x^i,\ \  
\sum_{i=0}^{\lfloor \frac{n}{2} \rfloor}\ f(\lfloor \frac{n}{2} \rfloor,i)\ x^i,\ \ 
\sum_{i=0}^{\lfloor \frac{q-n}{2} \rfloor}\ \binom{\lfloor \frac{q-n}{2} \rfloor}{i}\ x^i,\ \ 
\sum_{i=0}^{\lfloor \frac{q-n}{2} \rfloor}\ f(\lfloor \frac{q-n}{2} \rfloor,i)\ x^i\ ,
\ee
they do not lead to a desirable integrand and so are not included in table \ref{BuildingBlockSums}. 

In order to make the approach more transparent, let us consider the following example. Pick two individual sums:
\be
\sum_{i=0}^\infty\ \frac{\Gamma(i+d)}{\Gamma(i+1)\Gamma(d)} x_1^i \text{\ \ \ and\ \ \ } \sum_{j=0}^{i}\ \binom{i}{j}\ x_2^j\ ,
\ee
and choose a combination of allowed $[p_1,p_2]$ representing beta functions, for example $[2i,0]$ and $[j,0]$ (two since we are discussing triangle integrations), as prescribed in table \ref{BuildingBlockSums}. Concatenating the two chosen building blocks we obtain:
\ba
&&\sum_{i=0}^\infty\ \frac{\Gamma(i+d)}{\Gamma(i+1)\Gamma(d)} c_1^i \frac{\Gamma(2i+a_1+1)\Gamma(a_2+1)}{\Gamma(2i+a_{12}+2)} \sum_{j=0}^{i}\ \binom{i}{j}\ c_2^j \frac{\Gamma(j+a_3+1)\Gamma(a_4+1)}{\Gamma(j+a_{34}+2)}\nl
&=&\sum_{i=0}^\infty\ \frac{\Gamma(i+d)}{\Gamma(i+1)\Gamma(d)} c_1^i \sum_{j=0}^{i}\ \binom{i}{j}\ c_2^j \int_0^1\ du\ u_1^{2i+a_1} (1-u_1)^{a_2} u_2^{j+a_3} (1-u_2)^{a_4}\nl
&=&\int_0^1\ du\ u_1^{a_1} (1-u_1)^{a_2} u_2^{a_3} (1-u_2)^{a_4} \sum_{i=0}^\infty\ \frac{\Gamma(i+d)}{\Gamma(i+1)\Gamma(d)} \(c_1\ u_1^2\)^i \sum_{j=0}^{i}\ \binom{i}{j}\ \(c_2\ u_2\)^j\nl
&=&\int_0^1\ \frac{du\ u_1^{a_1} (1-u_1)^{a_2} u_2^{a_3} (1-u_2)^{a_4}}{\(1-c_1\ u_1^2\(1+c_2\ u_2\)\)^d}\,
\ea
which is a loop-like integration and a special case of eq. (\ref{TriangleScalar}). Note that since there are enough degrees of freedom ($c_1$ and $c_2$) the expansion is convergent. 

Going back to the building blocks presented in table \ref{BuildingBlockSums}, we now would like to consider every possible way to concatenate them. A very useful method is to visualize them using rooted trees as a guideline.\footnote{A related use of rooted trees and their relation to Hopf algebras can be found in Appendix A of \cite{Moch:2001zr}.} A rooted tree is composed of knots and links between knots. A connected rooted tree has a single root on top. A knot may be linked to several lower knots, but only one upper one. 

We will be considering rooted trees involving up to four knots, since this will be sufficient for the present discussion, however we note that rooted trees can be related to summations with any number of concatenated sums. We start by finding all connected rooted trees including up to four knots, and from these obtain all concatenated sums, as shown in table \ref{SumsConcatenation}.

\begin{table}[tp]
\begin{center}
\begin{tabular}{|c|l|}
\hline
\begin{picture}(10,10)(0,0)
\Vertex(4,4){1}
\end{picture} & $()$ \\ \hline \hline

\begin{picture}(10,15)(0,0)
\Vertex(5,10){1}
\Line(5,10)(5,0)
\Vertex(5,0){1}
\end{picture}
& $(i)$ $(2i)$\\ \hline \hline

\begin{picture}(10,15)(0,0)
\Vertex(5,10){1}
\Line(5,10)(0,0)
\Vertex(0,0){1}
\Line(5,10)(10,0)
\Vertex(10,0){1}
\end{picture}
&$(i,i)$ $(i,2i)$ $(2i,2i)$\\ \hline

\begin{picture}(10,25)(0,0)
\Vertex(5,20){1}
\Line(5,20)(5,10)
\Vertex(5,10){1}
\Line(5,10)(5,0)
\Vertex(5,0){1}
\end{picture}
&$(i,j)$ $(i,2j)$ $(2i,j)$ $(2i,2j)$\\ \hline\hline

\begin{picture}(22,15)(0,0)
\Vertex(10,10){1}
\Line(10,10)(20,0)
\Vertex(20,0){1}
\Line(10,10)(10,0)
\Vertex(10,0){1}
\Line(10,10)(0,0)
\Vertex(0,0){1}
\end{picture}
&$(i,i,i)$ $(i,i,2i)$ $(i,2i,2i)$ $(2i,2i,2i)$\\ \hline

\multirow{2}{*}{
\begin{picture}(12,25)(0,0)
\Vertex(5,20){1}
\Line(5,20)(0,10)
\Vertex(0,10){1}
\Line(5,20)(10,10)
\Vertex(10,10){1}
\Line(10,10)(10,0)
\Vertex(10,0){1}
\end{picture} }
& $(i,i,k)$ $(i,i,2k)$ $(i,2i,k)$ $(i,2i,2k)$ $(2i,i,k)$ $(2i,i,2k)$\\
& $(2i,2i,k)$ $(2i,2i,2k)$\\ \hline

\multirow{2}{*}{
\begin{picture}(12,25)(0,0)
\Vertex(5,20){1}
\Line(5,20)(5,10)
\Vertex(5,10){1}
\Line(5,10)(10,0)
\Vertex(10,0){1}
\Line(5,10)(0,0)
\Vertex(0,0){1}
\end{picture} }
& $(i,j,j)$ $(i,j,i-j)$ $(i,j,2j)$ $(i,j,2i-2j)$ $(2i,j,j)$ $(2i,j,2i-j)$\\
& $(2i,j,2j)$ $(2i,j,4i-2j)$\\ \hline

\multirow{2}{*}{
\begin{picture}(12,25)(0,0)
\Vertex(5,30){1}
\Line(5,30)(5,20)
\Vertex(5,20){1}
\Line(5,20)(5,10)
\Vertex(5,10){1}
\Line(5,10)(5,0)
\Vertex(5,0){1}
\end{picture} }
& \rule{0pt}{4ex}$(i,j,k)$ $(i,j,2k)$ $(i,2j,k)$ $(i,2j,2k)$ $(2i,j,k)$ $(2i,j,2k)$\\
& $(2i,2j,k)$ $(2i,2j,2k)$\\ \hline
\end{tabular}
\caption{All fully concatenated summations (up to four sums) using connected rooted trees as a guideline. The terms in parentheses represent the upper limit on summations from second to last, with the outer summation going to infinity. For example: $(i,2j,k) = \sum_{i=0}^\infty \sum_{j=0}^{i} \sum_{k=0}^{2j} \sum_{l=0}^k$. \label{SumsConcatenation}}
\end{center}
\end{table}

Next we need to also consider trees with disconnected branches, equivalent to summations involving more than one sum with infinity as the upper limit. This can be done by combining connected trees with the correct number of knots, as shown in table \ref{AllRootedTrees4}. 

{\renewcommand{\arraystretch}{1.3}
\renewcommand{\tabcolsep}{0.2cm}
\begin{table}[t]
\begin{center}
\begin{tabular}{|c|c c c c c c c c c|}
\hline
\#Sums & \multicolumn{9}{c|}{Rooted Trees} \\ \hline
1
&\begin{picture}(20,10)(0,0)
\Vertex(10,5){1}
\end{picture} & & & & & & & &\\ \hline

2
&\begin{picture}(20,10)(0,0)
\Vertex(7,7){1}
\Vertex(12,7){1}
\end{picture}
&\begin{picture}(20,10)(0,0)
\Vertex(10,6){1}
\Line(10,6)(10,0)
\Vertex(10,0){1}
\end{picture}& & & & & & & \\ \hline

3
&\begin{picture}(20,15)(0,2)
\Vertex(5,12){1}
\Vertex(10,12){1}
\Vertex(15,12){1}
\end{picture}
&\begin{picture}(20,15)(0,2)
\Vertex(7,12){1}
\Vertex(12,12){1}
\Line(12,12)(12,6)
\Vertex(12,6){1}
\end{picture}
&\begin{picture}(20,15)(0,2)
\Vertex(10,12){1}
\Line(10,12)(13,6)
\Vertex(13,6){1}
\Line(10,12)(7,6)
\Vertex(7,6){1}
\end{picture}
&\begin{picture}(20,15)(0,2)
\Vertex(10,12){1}
\Line(10,12)(10,6)
\Vertex(10,6){1}
\Line(10,6)(10,0)
\Vertex(10,0){1}
\end{picture} & & & & & \\ \hline

4
&\begin{picture}(20,18)(0,2)
\Vertex(3,18){1}
\Vertex(8,18){1}
\Vertex(13,18){1}
\Vertex(18,18){1}
\end{picture}
&\begin{picture}(20,18)(0,2)
\Vertex(5,18){1}
\Vertex(10,18){1}
\Vertex(15,18){1}
\Line(15,18)(15,12)
\Vertex(15,12){1}
\end{picture}
&\begin{picture}(20,18)(0,2)
\Vertex(7,18){1}
\Vertex(13,18){1}
\Line(13,18)(10,12)
\Vertex(10,12){1}
\Line(13,18)(16,12)
\Vertex(16,12){1}
\end{picture}
&\begin{picture}(20,20)(0,2)
\Vertex(7,18){1}
\Vertex(12,18){1}
\Line(12,18)(12,12)
\Vertex(12,12){1}
\Line(12,12)(12,6)
\Vertex(12,6){1}
\end{picture}
&\begin{picture}(20,18)(0,2)
\Vertex(7,18){1}
\Vertex(12,18){1}
\Line(7,18)(7,12)
\Vertex(7,12){1}
\Line(12,18)(12,12)
\Vertex(12,12){1}
\end{picture}
&\begin{picture}(20,18)(0,2)
\Vertex(10,18){1}
\Line(10,18)(10,12)
\Vertex(10,12){1}
\Line(10,18)(15,12)
\Vertex(15,12){1}
\Line(10,18)(5,12)
\Vertex(5,12){1}
\end{picture}
&\begin{picture}(20,18)(0,2)
\Vertex(10,18){1}
\Line(10,18)(7,12)
\Vertex(7,12){1}
\Line(10,18)(13,12)
\Vertex(13,12){1}
\Line(13,12)(13,6)
\Vertex(13,6){1}
\end{picture}
&\begin{picture}(20,18)(0,2)
\Vertex(10,18){1}
\Line(10,18)(10,12)
\Vertex(10,12){1}
\Line(10,12)(7,6)
\Vertex(7,6){1}
\Line(10,12)(13,6)
\Vertex(13,6){1}
\end{picture}
&\begin{picture}(20,18)(0,2)
\Vertex(10,18){1}
\Line(10,18)(10,12)
\Vertex(10,12){1}
\Line(10,12)(10,6)
\Vertex(10,6){1}
\Line(10,6)(10,0)
\Vertex(10,0){1}
\end{picture} \\ \hline

\end{tabular}
\caption{All rooted trees up to four knots built by joining connected rooted trees. \label{AllRootedTrees4}}
\end{center}
\end{table}
}

Following this procedure, we obtain the results presented in table \ref{SeamlessPolynomialsByReverseMethod}. These results may be used to compare with the parameter integration obtained from specific diagrams.

{\setlength{\tabcolsep}{2pt}
\setlength{\LTcapwidth}{14cm}
\begin{center}
\begin{longtable}{c l l}
\caption{Parameter integrations leading to summations with \emph{seamless reduction}. The integration form is given by:
$\int \frac{du_1\ du_2\ u_1^{a_1-1}\ (1-u_1)^{a_2-1}\ u_2^{a_3-1}\ (1-u_2)^{a_4-1}}{D^d}$, where $D$ is the denominator presented in each row and $a_\ell$ are constants.
$\sum _{i_\ell}^{k}$ represents $\sum_{i_\ell=0}^k c_\ell^{i_\ell}$, where $c_\ell$ is a given coefficient, $d_i=\frac{\Gamma(i+d)}{\Gamma(d)\Gamma(i+1)}$, $b_i^j=\frac{\Gamma(i+a_{(s)}) \Gamma(j+a_{(s+1)})}{\Gamma(i+j+a_{(s)}+a_{(s+1)})}$, and $b_i=b_i^0$, with $s$ equal to $1$ or $3$.} \label{SeamlessPolynomialsByReverseMethod} \\

\hline 
\# & Denominator ($D$)& Sum \\ \hline
\endfirsthead

\multicolumn{3}{c}{\tablename\ \thetable{} continued} \\
\hline 
\# & Denominator ($D$)& Sum \\ \hline
\endhead

\hline \multicolumn{2}{c}{Continued on next page} \\ 
\endfoot

\hline
\endlastfoot

1 & {\small $\left(1-c_1 u_2\right) $} & $\sum _{i_1}^{\infty } d_{i_1} b_{0} b_{i_1}  $ \\ \hline

2 & {\small $\left(1-c_1 u_2^2\right)$ } & $\sum _{i_1}^{\infty } d_{i_1} b_{0} b_{2 i_1}  $ \\ \hline

3 & {\small $\left(1-c_1 u_1 u_2\right)$ } & $\sum _{i_1}^{\infty } d_{i_1} b_{i_1} b_{i_1}  $ \\ \hline

4 & {\small $\left(1-c_1 u_1 u_2^2\right)$ } & $\sum _{i_1}^{\infty } d_{i_1} b_{i_1} b_{2 i_1}  $ \\ \hline

5 & {\small $\left(1-c_1 u_1^2 u_2^2\right)$ } & $\sum _{i_1}^{\infty } d_{i_1} b_{2 i_1} b_{2 i_1}  $ \\ \hline

6 & {\small $\left(\left(1-c_1 u_1\right) \left(1-c_2 u_2\right)\right)$ } & $\sum _{i_1}^{\infty } \sum _{i_2}^{\infty } d_{i_1} d_{i_2} b_{i_1} b_{i_2}   $ \\ \hline 

7 & {\small $\left(\left(1-c_1 u_1\right) \left(1-c_2 u_2^2\right)\right)$ } & $\sum _{i_1}^{\infty } \sum _{i_2}^{\infty } d_{i_1} d_{i_2} b_{i_1} b_{2 i_2}   $ \\ \hline 

8 & {\small $\left(\left(1-c_1 u_1^2\right) \left(1-c_2 u_2^2\right)\right)$ } & $\sum _{i_1}^{\infty } \sum _{i_2}^{\infty } d_{i_1} d_{i_2} b_{2 i_1} b_{2 i_2}   $ \\ \hline 

9 & {\small $\left(\left(1-c_1 u_2\right) \left(1-c_1 c_2 u_2\right)\right)$ } & $\sum _{i_1}^{\infty } \sum _{i_2}^{i_1} d_{i_1-i_2} d_{i_2} b_{0} b_{i_1}   $ \\ \hline 

10 & {\small $\left(\left(1-c_1 u_1 u_2\right) \left(1-c_1 c_2 u_1 u_2\right)\right)$ } & $\sum _{i_1}^{\infty } \sum _{i_2}^{i_1} d_{i_1-i_2} d_{i_2} b_{i_1} b_{i_1}   $ \\ \hline 

11 & {\small $\left(\left(1-c_1 u_1\right) \left(1-c_1 c_2 u_1 u_2\right)\right)$ } & $\sum _{i_1}^{\infty } \sum _{i_2}^{i_1} d_{i_1-i_2} d_{i_2} b_{i_1} b_{i_2}   $ \\ \hline 

12 & {\small $\left(\left(1-c_1 u_1\right) \left(1-c_1 c_2 u_1 u_2^2\right)\right)$ } & $\sum _{i_1}^{\infty } \sum _{i_2}^{i_1} d_{i_1-i_2} d_{i_2} b_{i_1} b_{2 i_2}   $ \\ \hline 

13 & {\small $\left(1-c_1 \left(1+\frac{c_2 u_1}{u_2}\right) u_2\right)$ } & $\sum _{i_1}^{\infty } \sum _{i_2}^{i_1} \binom{i_1}{i_2} d_{i_1} b_{i_1-i_2} b_{i_2}   $ \\ \hline 

14 & {\small $\left(1-c_1 \left(1+\frac{c_2 u_1 \left(1-u_2\right)}{u_2}\right) u_2\right)$ } & $\sum _{i_1}^{\infty } \sum _{i_2}^{i_1} \binom{i_1}{i_2} d_{i_1} b_{i_1-i_2}^{i_2} b_{i_2}   $ \\ \hline 

15 & {\small $\left(1-c_1 \left(1+c_2 u_1\right) u_2\right)$ } & $\sum _{i_1}^{\infty } \sum _{i_2}^{i_1} \binom{i_1}{i_2} d_{i_1} b_{i_1} b_{i_2}   $ \\ \hline 

16 & {\small $\left(1-c_1 \left(1+c_2 u_1\right) u_2^2\right)$ } & $\sum _{i_1}^{\infty } \sum _{i_2}^{i_1} \binom{i_1}{i_2} d_{i_1} b_{2 i_1} b_{i_2}   $ \\ \hline 

17 & {\small $\left(1-c_1 u_1 \left(1+\frac{c_2 \left(1-u_1\right) u_2^2}{u_1}\right)\right)$ } & $\sum _{i_1}^{\infty } \sum _{i_2}^{i_1} \binom{i_1}{i_2} d_{i_1} b_{i_1-i_2,i_2} b_{2 i_2}   $ \\ \hline 

18 & {\small $\left(1-c_1 u_1 \left(1+\frac{c_2 \left(1-u_1\right)}{u_1 u_2^2}\right) u_2^2\right)$ } & $\sum _{i_1}^{\infty } \sum _{i_2}^{i_1} \binom{i_1}{i_2} d_{i_1} b_{2 i_1-2 i_2} b_{i_1-i_2}^{i_2}   $ \\ \hline 

19 & {\small $\left(1-c_1 \left(1+\frac{c_2 u_1}{u_2}\right)^2 u_2^2\right)$ } & $\sum _{i_1}^{\infty } \sum _{i_2}^{2 i_1}\binom{2 i_1}{i_2}  d_{i_1} b_{2 i_1-i_2} b_{i_2}   $ \\ \hline 

20 & {\small $\left(1-c_1 \left(1+\frac{c_2 u_1 \left(1-u_2\right)}{u_2}\right)^2 u_2^2\right)$ } & $\sum _{i_1}^{\infty } \sum _{i_2}^{2 i_1} \binom{2 i_1}{i_2} d_{i_1} b_{2 i_1-i_2}^{i_2} b_{i_2}   $ \\ \hline 

21 & {\small $\left(1-c_1 \left(1+c_2 u_1\right)^2 u_2\right)$ } & $\sum _{i_1}^{\infty } \sum _{i_2}^{2 i_1} \binom{2 i_1}{i_2} d_{i_1} b_{i_1} b_{i_2}   $ \\ \hline 

22 & {\small $\left(1-c_1 \left(1+c_2 u_1\right)^2 u_2^2\right)$ } & $\sum _{i_1}^{\infty } \sum _{i_2}^{2 i_1} \binom{2 i_1}{i_2} d_{i_1} b_{2 i_1} b_{i_2}   $ \\ \hline 

23 & {\small $\left(1-c_1 u_1^2 \left(1+\frac{c_2 \left(1-u_2\right)}{u_1 u_2}\right)^2 u_2^2\right)$ } & $\sum _{i_1}^{\infty } \sum _{i_2}^{2 i_1} \binom{2 i_1}{i_2} d_{i_1} b_{2 i_1-i_2} b_{2 i_1-i_2}^{i_2}   $ \\ \hline 

24 & {\small $\left(\left(1-c_1 u_1\right) \left(1-c_1 c_3 u_1\right) \left(1-c_2 u_2\right)\right)$ } & $\sum _{i_1}^{\infty } \sum _{i_2}^{\infty } \sum _{i_3}^{i_1} d_{i_2} d_{i_1-i_3} d_{i_3} b_{i_1} b_{i_2}    $ \\ \hline 

25 & {\small $\left(\left(1-c_1 u_1\right) \left(1-c_1 c_3 u_1\right) \left(1-c_2 u_2^2\right)\right)$ } & $\sum _{i_1}^{\infty } \sum _{i_2}^{\infty } \sum _{i_3}^{i_1} d_{i_2} d_{i_1-i_3} d_{i_3} b_{i_1} b_{2 i_2}    $ \\ \hline 

26 & {\small $\left(1-c_1 \left(1+c_2 u_1\right) \left(1+c_3 u_2\right)\right)$ } & $\sum _{i_1}^{\infty } \sum _{i_2}^{i_1} \sum _{i_3}^{i_1} \binom{i_1}{i_2} \binom{i_1}{i_3} d_{i_1} b_{i_2} b_{i_3}    $ \\ \hline 

27 & {\small $\left(1-c_1 \left(1+c_2 u_1\right) \left(1+c_3 u_2\right)^2\right)$ } & $\sum _{i_1}^{\infty } \sum _{i_2}^{i_1} \sum _{i_3}^{2 i_1} \binom{i_1}{i_2} \binom{2 i_1}{i_3} d_{i_1} b_{i_2} b_{i_3}    $ \\ \hline 

28 & {\small $\left(1-c_1 \left(1+c_2 u_1\right)^2 \left(1+c_3 u_2\right)^2\right)$ } & $\sum _{i_1}^{\infty } \sum _{i_2}^{2 i_1} \sum _{i_3}^{2 i_1} \binom{2 i_1}{i_2} \binom{2 i_1}{i_3} d_{i_1} b_{i_2} b_{i_3}    $ \\ \hline 

29 & {\small $\left(\left(1-c_1 u_2\right) \left(1-c_1 c_2 \left(1+c_3 u_1\right) u_2\right)\right)$ } & $\sum _{i_1}^{\infty } \sum _{i_2}^{i_1} \sum _{i_3}^{i_2} \binom{i_2}{i_3} d_{i_1-i_2} d_{i_2} b_{i_1} b_{i_3}    $ \\ \hline 

30 & {\small $\left(\left(1-c_1 c_2 u_2\right) \left(1-c_1 \left(1+\frac{c_3}{u_1}\right) u_1 u_2\right)\right)$ } & $\sum _{i_1}^{\infty } \sum _{i_2}^{i_1} \sum _{i_3}^{i_1-i_2} \binom{i_1-i_2}{i_3} d_{i_1-i_2} d_{i_2} b_{i_1} b_{i_1-i_2-i_3}$ \\ \hline 

31 & {\small $\left(\left(1-c_1 u_2\right) \left(1-c_1 c_2 \left(1+c_3 u_1\right)^2 u_2\right)\right)$ } & $\sum _{i_1}^{\infty } \sum _{i_2}^{i_1} \sum _{i_3}^{2 i_2} \binom{2 i_2}{i_3} d_{i_1-i_2} d_{i_2} b_{i_1} b_{i_3}$ \\ \hline 

32 & {\small $\left(1-c_1 \left(1+c_2 u_1 \left(1+c_3 u_2\right)\right)^2\right)$ } & $\sum _{i_1}^{\infty } \sum _{i_2}^{2 i_1} \sum _{i_3}^{i_2} \binom{2 i_1}{i_2} \binom{i_2}{i_3} d_{i_1} b_{i_2} b_{i_3}    $ \\ \hline 

33 & {\small $\left(1-c_1 u_1^2 \left(1+\frac{c_2 \left(1+c_3 u_2\right)}{u_1}\right)^2\right)$ } & $\sum _{i_1}^{\infty } \sum _{i_2}^{2 i_1} \sum _{i_3}^{i_2}  \binom{2 i_1}{i_2} \binom{i_2}{i_3} d_{i_1} b_{2 i_1-i_2} b_{i_3}    $ \\ \hline 

34 & {\small $\left(1-c_1 u_1^2 \left(1+\frac{c_2 \left(1-u_1\right) \left(1+\frac{c_3}{u_2}\right) u_2}{u_1}\right)^2\right)$ } & $\sum _{i_1}^{\infty } \sum _{i_2}^{2 i_1} \sum _{i_3}^{i_2} \binom{2 i_1}{i_2} \binom{i_2}{i_3} d_{i_1} b_{2 i_1-i_2}^{i_2} b_{i_2-i_3}$ \\ \hline 

35 & {\small $\left(1\!-\!c_1 u_1\right) \left(1\!-\!c_1 c_3 u_1\right) \left(1\!-\!c_2 u_2\right) \left(1\!-\!c_2 c_4 u_2\right)$ }&$\sum _{i_1}^{\infty } \sum _{i_2}^{\infty } \sum _{i_3}^{i_1} \sum _{i_4}^{i_2} d_{i_1-i_3} d_{i_3} d_{i_2-i_4} d_{i_4} b_{i_1} b_{i_2} $ \\ \hline 

36 & {\small $\left(1-c_1 \left(1+c_2 \left(1+c_3 u_1\right) \left(1+c_4 u_2\right)\right)^2\right)$ }&$\sum _{i_1}^{\infty } \sum _{i_2}^{2 i_1} \sum _{i_3}^{i_2} \sum _{i_4}^{i_2} \binom{2 i_1}{i_2} \binom{i_2}{i_3} \binom{i_2}{i_4} d_{i_1} b_{i_3} b_{i_4} $ \\ \hline 

37 & \multicolumn{2}{l}{{\small $\left(1\!-\!c_1 c_2  u_2\right) \left(1\!-\!c_1 \left(1\!+\!c_4 u_1\right) \left(1\!+\!\frac{c_3}{1\!+\!c_4 u_1}\right) u_2\right)$ }} \\*
&\multicolumn{2}{r}{$\sum _{i_1}^{\infty } \sum _{i_2}^{i_1} \sum _{i_3}^{i_1-i_2} \sum _{i_4}^{i_1-i_2-i_3} \binom{i_1-i_2}{i_3} \binom{i_1-i_2-i_3}{i_4} d_{i_1-i_2} d_{i_2} b_{i_1} b_{i_4}$} \\ \hline 
\end{longtable}
\end{center}
}

The denominators presented in table \ref{SeamlessPolynomialsByReverseMethod} are the result of applying the backward approach with two beta functions, and have been obtained with a software package we developed using Mathematica. They represent all loop-like integrations involving three propagators that may be reduced seamlessly using the available Z-Sum algorithms A, B, C, and D (eqs. (\ref{AlgA}-\ref{AlgD})), although not all will necessarily correspond to a loop diagram. No new solutions are found involving more than four sums.

\subsection{Survey of triangles in QFT}
Now that we known which loop-like integrals can be seamlessly reduced, we would like to cross check this result with actual loop calculations. In order to consider every possible triangle in QFT we assign all possible values to the six invariants $(p_1^2,p_2^2,p_3^2,m_1^2,m_2^2,m_3^2)$, where $p_\ell$ represents the external momenta and $m$ are the masses of internal particles. If we consider only loops with at least one external leg off-shell, this gives a total of 101 triangles, out of which 72 are present in the Standard Model, as shown in table \ref{TriaglesQFT-SM}.

\begin{table}[tp]
\begin{center}
{\small
\begin{tabular}{|ccc|ccc|ccc|}
\hline
\#  &{\tiny$(p_1^2,p_2^2,p_3^2,m_1^2,m_2^2,m_3^2)$} &S&\#  & {\tiny$(p_1^2,p_2^2,p_3^2,m_1^2,m_2^2,m_3^2)$}&S&\# & {\tiny$(p_1^2,p_2^2,p_3^2,m_1^2,m_2^2,m_3^2)$}&S\\ \hline
1  &$\color{red}{[0,0,c_1,0,0,0]}$ 		&0 &2  & $(0,0,c_1,0,0,c_2)$ 			  &14&3 & $\color{red}{[0,0,c_1,0,c_2,c_2]}$ 		& \\
4  &$\color{red}{[0,0,c_1,c_2,0,0]}$ 		&14&5  & $(0,0,c_1,c_2,0,c_2)$ 			  &14&6 & $\color{red}{[0,0,c_1,c_2,c_2,c_2]}$ 		& \\
7  &$\color{red}{[0,c_1,c_2,0,0,0]}$ 		&15&8  & $\color{red}{[0,c_1,c_2,0,0,c_1]}$ 	  &26&9 & $(0,c_1,c_2,0,0,c_2)$ 			&26\\
10 &$\color{red}{[0,c_1,c_2,0,c_1,0]}$		&15&11 & $(0,c_1,c_2,0,c_1,c_1)$ 		  &  &12& $\color{red}{[0,c_1,c_2,0,c_2,c_2]}$ 		& \\
13 &$\color{red}{[0,c_1,c_2,c_1,0,c_1]}$ 	&14&14 & $\color{red}{[0,c_1,c_2,c_1,c_1,c_1]}$   &  &15& $\color{red}{[c_1,c_1,c_2,0,0,0]}$ 		& \\
16 &$(c_1,c_1,c_2,0,0,c_1)$ 			&  &17 & $\color{red}{[c_1,c_1,c_2,0,c_1,c_1]}$   &9 &18& $\color{red}{[c_1,c_1,c_2,c_1,0,0]}$		& \\
19 &$\color{red}{[c_1,c_1,c_2,c_1,0,c_1]}$	&  &20 & $\color{red}{[c_1,c_1,c_2,c_1,c_1,c_1]}$ &  &21& $\color{red}{[0,0,c_1,0,c_2,c_3]}$ 		& \\
22 &$(0,0,c_1,c_2,0,c_3)$ 			&14&23 & $\color{red}{[0,0,c_1,c_2,c_2,c_3]}$ 	  &  &24& $\color{red}{[0,0,c_1,c_2,c_3,c_3]}$ 		& \\ 
25 &$\color{red}{[0,c_1,c_2,0,0,c_3]}$ 		&26&26 & $(0,c_1,c_2,0,c_1,c_3)$ 		  &  &27& $\color{red}{[0,c_1,c_2,0,c_2,c_3]}$ 		& \\ 
28 &$\color{red}{[0,c_1,c_2,0,c_3,0]}$ 		&15&29 & $(0,c_1,c_2,0,c_3,c_1)$ 		  &  &30& $\color{red}{[0,c_1,c_2,0,c_3,c_2]}$ 		& \\ 
31 &$\color{red}{[0,c_1,c_2,0,c_3,c_3]}$ 	&  &32 & $\color{red}{[0,c_1,c_2,c_1,0,c_3]}$ 	  &26&33& $\color{red}{[0,c_1,c_2,c_1,c_1,c_3]}$ 	& \\ 
34 &$\color{red}{[0,c_1,c_2,c_1,c_3,c_1]}$	&  &35 & $(0,c_1,c_2,c_1,c_3,c_3)$ 		  &  &36& $(0,c_1,c_2,c_2,0,c_3)$ 			&26\\ 
37 &$(0,c_1,c_2,c_2,c_2,c_3)$ 			&  &38 & $(0,c_1,c_2,c_2,c_3,c_3)$ 		  &  &39& $\color{red}{[0,c_1,c_2,c_3,0,c_3]}$ 		&14\\
40 &$(0,c_1,c_2,c_3,c_1,c_3)$ 			&  &41 & $\color{red}{[0,c_1,c_2,c_3,c_3,c_3]}$   &  &42& $(c_1,c_1,c_2,0,0,c_3)$ 			& \\
43 &$(c_1,c_1,c_2,0,c_1,c_3)$ 			&  &44 & $\color{red}{[c_1,c_1,c_2,0,c_3,c_3]}$   &  &45& $(c_1,c_1,c_2,c_1,0,c_3)$			& \\
46 &$\color{red}{[c_1,c_1,c_2,c_1,c_1,c_3]}$	&  &47 & $\color{red}{[c_1,c_1,c_2,c_1,c_3,c_3]}$ &  &48& $\color{red}{[c_1,c_1,c_2,c_3,0,0]}$ 		& \\
49 &$(c_1,c_1,c_2,c_3,0,c_1)$ 			&  &50 & $(c_1,c_1,c_2,c_3,0,c_3)$ 		  &  &51& $\color{red}{[c_1,c_1,c_2,c_3,c_1,c_1]}$	& \\
52 &$(c_1,c_1,c_2,c_3,c_1,c_3)$ 		&  &53 & $\color{red}{[c_1,c_1,c_2,c_3,c_3,c_3]}$ &  &54& $\color{red}{[c_1,c_2,c_3,0,0,0]}$ 		& \\
55 &$\color{red}{[c_1,c_2,c_3,0,0,c_1]}$ 	&  &56 & $\color{red}{[c_1,c_2,c_3,0,0,c_2]}$ 	  &  &57& $\color{red}{[c_1,c_2,c_3,0,c_1,c_1]}$ 	& \\
58 &$(c_1,c_2,c_3,0,c_1,c_2)$ 			&  &59 & $(c_1,c_2,c_3,0,c_1,c_3)$ 		  &  &60& $\color{red}{[c_1,c_2,c_3,0,c_2,c_1]}$ 	&9\\
61 &$(c_1,c_2,c_3,0,c_2,c_3)$ 			&  &62 & $\color{red}{[c_1,c_2,c_3,0,c_3,c_3]}$   &  &63& $\color{red}{[c_1,c_2,c_3,c_1,c_1,c_1]}$ 	& \\
64 &$\color{red}{[c_1,c_2,c_3,c_1,c_1,c_2]}$	&  &65 & $(c_1,c_2,c_3,c_1,c_1,c_3)$ 		  &  &66& $\color{red}{[c_1,c_2,c_3,c_1,c_2,c_1]}$ 	& \\
67 &$(0,0,c_1,c_2,c_3,c_4)$ 			&  &68 & $\color{red}{[0,c_1,c_2,0,c_3,c_4]}$ 	  &  &69& $\color{red}{[0,c_1,c_2,c_1,c_3,c_4]}$ 	& \\
70 &$(0,c_1,c_2,c_2,c_3,c_4)$ 			&  &71 & $\color{red}{[0,c_1,c_2,c_3,0,c_4]}$ 	  &26&72& $(0,c_1,c_2,c_3,c_1,c_4)$ 			& \\
73 &$\color{red}{[0,c_1,c_2,c_3,c_3,c_4]}$	&  &74 & $\color{red}{[0,c_1,c_2,c_3,c_4,c_3]}$   &  &75& $(c_1,c_1,c_2,0,c_3,c_4)$ 			& \\
76 &$\color{red}{[c_1,c_1,c_2,c_1,c_3,c_4]}$	&  &77 & $(c_1,c_1,c_2,c_3,0,c_4)$ 		  &  &78& $\color{red}{[c_1,c_1,c_2,c_3,c_1,c_4]}$ 	& \\
79 &$\color{red}{[c_1,c_1,c_2,c_3,c_3,c_4]}$	&  &80 & $\color{red}{[c_1,c_1,c_2,c_3,c_4,c_4]}$ &  &81& $\color{red}{[c_1,c_2,c_3,0,0,c_4]}$ 		& \\
82 &$(c_1,c_2,c_3,0,c_1,c_4)$ 			&  &83 & $\color{red}{[c_1,c_2,c_3,0,c_2,c_4]}$   &  &84& $\color{red}{[c_1,c_2,c_3,0,c_3,c_4]}$ 	& \\
85 &$\color{red}{[c_1,c_2,c_3,0,c_4,c_4]}$	&  &86 & $\color{red}{[c_1,c_2,c_3,c_1,c_1,c_4]}$ &  &87& $\color{red}{[c_1,c_2,c_3,c_1,c_2,c_4]}$ 	& \\
88 &$\color{red}{[c_1,c_2,c_3,c_1,c_3,c_4]}$	&  &89 & $\color{red}{[c_1,c_2,c_3,c_1,c_4,c_1]}$ &  &90& $\color{red}{[c_1,c_2,c_3,c_1,c_4,c_2]}$ 	& \\
91 &$\color{red}{[c_1,c_2,c_3,c_1,c_4,c_4]}$	&  &92 & $\color{red}{[c_1,c_2,c_3,c_3,c_1,c_4]}$ &  &93& $\color{red}{[c_1,c_2,c_3,c_3,c_4,c_4]}$ 	& \\
94 &$\color{red}{[c_1,c_2,c_3,c_4,c_4,c_4]}$	&  &95 & $\color{red}{[0,c_1,c_2,c_3,c_4,c_5]}$   &  &96& $\color{red}{[c_1,c_1,c_2,c_3,c_4,c_5]}$ 	& \\
97 &$\color{red}{[c_1,c_2,c_3,0,c_4,c_5]}$ 	&  &98 & $\color{red}{[c_1,c_2,c_3,c_1,c_4,c_5]}$ &  &99& $\color{red}{[c_1,c_2,c_3,c_3,c_4,c_5]}$ 	& \\
100&$\color{red}{[c_1,c_2,c_3,c_4,c_4,c_5]}$	&  &101& $\color{red}{[c_1,c_2,c_3,c_4,c_5,c_6]}$ &  &  & 						& \\ \hline
\end{tabular}
}
\caption{One-loop triangle diagrams in QFT ordered by degrees of freedom, where duplicates due to rotation or reflection symmetries have been excluded. Bracket refer to the invariants $(p_1^2,p_2^2,p_3^2,m_1^2,m_2^2,m_3^2)$, the external momenta and internal masses. Square brackets (red) represent cases present in the Standard Model. The third subcolumn (labeled S) shows diagrams with \emph{seamless reduction}, with the number indicating the corresponding polynomial in table \ref{SeamlessPolynomialsByReverseMethod}, and 0 indicating no expansion necessary. \label{TriaglesQFT-SM}}
\end{center}
\end{table}

The results presented in table \ref{TriaglesQFT-SM} show that only a fraction of the total number of triangles (18) can be reduced to Z-Sums using Taylor expansions and the reduction algorithms currently known (eqs. (\ref{AlgA}-\ref{AlgD})). All other diagrams lead to well defined expansions that cannot be systematically reduced to Z-Sums without solutions being found for the missing algorithms (eqs. (\ref{MA1}-\ref{MA4})). 

For some calculations where the Z-Sum approach with Taylor expansion could not be successfully applied, results may be known by other integration methods. These results can be related at the summation level, however the steps required for making the connection are not systematic, changing for every diagram and every order in $\eps$, and so are not considered seamless reductions because they defeat the purpose of developing a systematic approach.

\subsection{Boxes and beyond}
The procedure presented in this section can be directly applied to parameter integrations originating from momentum integrations with more propagators, like boxes or multi-loop integrations. The difference is that they would involve a larger number of parameter-integration variables, and so the denominator would potentially involve a larger number of terms. Both the forward and the backward approaches are applicable, where in the latter we would use a larger number of beta functions in order to find the results equivalent to table \ref{SeamlessPolynomialsByReverseMethod}. While we could proceed with this calculation, we do not believe it would be very insightful, since we expect similar results to those obtained for triangle loops. Specifically, we believe some of the calculations would be feasible using the existing algorithms A, B, C, and D (eqs. (\ref{AlgA}-\ref{AlgD})), while other would require a solution for the missing algorithms discussed previously (eqs. (\ref{MA1}-\ref{MA4})).

\section{Application}
\label{Application}

In this section we would like to focus on the successes of the approach by applying it to a problem of physical interest, and for the first time present results for multi-loop calculations involving massive internal particles using the Z-Sum reduction algorithms and Taylor expansions. One interesting application is the calculation of the heavy-flavor corrections to Deep Inelastic Scattering (DIS) structure functions, which can be relevant (up to order 20-40\%) for small values of the Bjorken variable $x$ \cite{Buza:1995ie,Blumlein:2006mh,Bierenbaum:2006mq,Bierenbaum:2007dm,Bierenbaum:2007pn,Bierenbaum:2007qe,Bierenbaum:2008yu,Bierenbaum:2009zt,Bierenbaum:2010jp}.

The scattering amplitude can be represented on a basis of matrix elements of universal operators whose coefficients are process dependent and can be calculated perturbatively in QCD. 
The effects of massive quarks factorize entirely in the operator matrix elements \cite{Buza:1995ie} and has been calculated analytically in the asymptotic region $Q^2 \gg m^2$ at  $O(\alpha_s^2)$ 
 \cite{Bierenbaum:2006mq,Bierenbaum:2007dm,Bierenbaum:2007pn,Bierenbaum:2007qe,Bierenbaum:2008yu,Bierenbaum:2009zt}, while partial contributions have been obtained at  $O(\alpha_s^3)$ \cite{Bierenbaum:2010jp}. In both cases, the effects are sizable and definitely to be included in the theoretical predictions to be safely compared with the current experimental results. These calculations include the most important set of contributions from operators with external heavy-flavor and gluon lines and have been performed using inverse Mellin-Barnes transformations (approach C, table \ref{TableABC}). As a test of the applicability and potential of the Taylor expansion method we have reproduced the results presented in \cite{Bierenbaum:2006mq}.

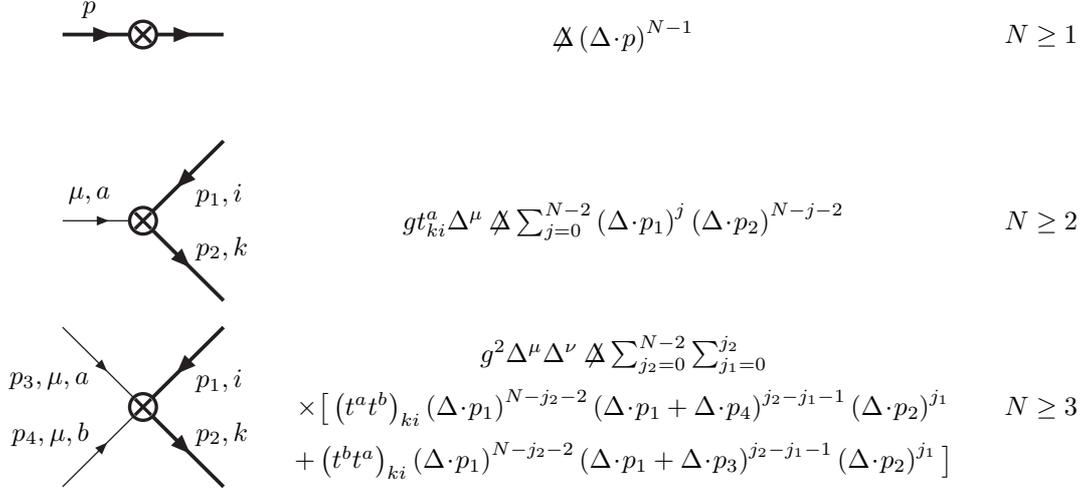
\begin{figure}[tp]
\begin{center}
\begin{picture}(400,210)(0,20)

{\SetWidth{1.4}
\ArrowLine(020,170)(050,170)
\ArrowLine(050,170)(080,170)
}
{\SetWidth{1.25}
\CCirc(50,170){5}{Black}{White}
\Line(47,167)(53,173)
\Line(47,173)(53,167)
}

\small{
\Text(030,180)[c]{$p$}
}

\Text(230,170)[c]{$\not\!\!\Delta\left(\Delta\!\cdot\!p\right)^{N-1}$}
\Text(400,170)[r]{$N \ge 1$}


\ArrowLine(020,100)(050,100)
{\SetWidth{1.4}
\ArrowLine(050,100)(080,070)
\ArrowLine(080,130)(050,100)
}
{\SetWidth{1.25}
\CCirc(50,100){5}{Black}{White}
\Line(47,97)(53,103)
\Line(47,103)(53,97)
}

\small{
\Text(030,110)[c]{$\mu, a$}
\Text(070,110)[l]{$p_1, i$}
\Text(070,090)[l]{$p_2, k$}
}

\Text(230,100)[c]{$g t_{ki}^a \Delta^\mu \not\!\!\Delta \sum_{j=0}^{N-2} \left(\Delta\!\cdot\!p_1\right)^{j} \left(\Delta\!\cdot\!p_2\right)^{N-j-2}$}
\Text(400,100)[r]{$N \ge 2$}


{\SetWidth{1.4}
\ArrowLine(050,030)(080,000)
\ArrowLine(080,060)(050,030)
}
\ArrowLine(020,000)(050,030)
\ArrowLine(020,060)(050,030)
{\SetWidth{1.25}
\CCirc(50,30){5}{Black}{White}
\Line(47,27)(53,33)
\Line(47,33)(53,27)
}

\small{
\Text(070,040)[l]{$p_1, i$}
\Text(070,020)[l]{$p_2, k$}
\Text(030,040)[r]{$p_3, \mu, a$}
\Text(030,020)[r]{$p_4, \mu, b$}
}

\Text(230,050)[c]{$g^2 \Delta^\mu \Delta^\nu \not\!\!\Delta \sum_{j_2=0}^{N-2} \sum_{j_1=0}^{j_2}$}
\Text(230,030)[c]{$\times \big[\left(t^a t^b\right)_{ki} \left(\Delta\!\cdot\!p_1\right)^{N-j_2-2} \left( \Delta\!\cdot\!p_1+ \Delta\!\cdot\!p_4 \right)^{j_2-j_1-1} \left(\Delta\!\cdot\!p_2\right)^{j_1}$}
\Text(230,010)[c]{$+\left(t^b t^a\right)_{ki} \left(\Delta\!\cdot\!p_1\right)^{N-j_2-2} \left( \Delta\!\cdot\!p_1+ \Delta\!\cdot\!p_3 \right)^{j_2-j_1-1} \left(\Delta\!\cdot\!p_2\right)^{j_1} \big]$}

\Text(400,030)[r]{$N \ge 3$}

\end{picture} 
\end{center}
\caption{\label{InsertionFeynmanRules} Feynman rules for operator insertions, where $\Delta$ is a light-like vector and N is the Mellin-space variable.}
\end{figure}

\begin{figure}[tp]
\begin{center}
\begin{picture}(400,120)(0,20)

\Text(100,120)[c]{\Large{A}}
\ArrowLine(010,050)(040,050)
{\SetWidth{1.75} \ArrowLine(040,050)(100,100)}
{\SetWidth{1.75} \ArrowLine(100,000)(040,050)}
\ArrowLine(100,100)(100,000)
{\SetWidth{1.75} \ArrowLine(160,050)(100,100)}
{\SetWidth{1.75} \ArrowLine(100,000)(160,050)}
\ArrowLine(160,050)(190,050)
\Text(010,060)[l]{$p$}
\Text(065,080)[r]{$k_1$}
\Text(135,080)[l]{$k_2$}
\Text(065,020)[r]{$k_1\!-\!p$}
\Text(135,020)[l]{$k_2\!+\!p$}
\Text(105,050)[l]{$k_1\!+\!k_2$}

\Text(300,120)[c]{\Large{B}}
\ArrowLine(210,050)(240,050)
{\SetWidth{1.75} \ArrowLine(240,050)(300,100)}
{\SetWidth{1.75} \ArrowLine(300,000)(240,050)}
{\SetWidth{1.75} \ArrowLine(300,100)(300,000)}
\ArrowLine(360,050)(300,100)
\ArrowLine(300,000)(360,050)
\ArrowLine(360,050)(390,050)
\Text(210,060)[l]{$p$}
\Text(265,080)[r]{$k_1$}
\Text(335,080)[l]{$k_2$}
\Text(265,020)[r]{$k_1\!-\!p$}
\Text(335,020)[l]{$k_2\!+\!p$}
\Text(305,050)[l]{$k_1\!+\!k_2$}
\end{picture} 
\end{center}
\caption{\label{DiagramsWithoutInsertion} Two-loop diagrams with massive internal particles (represented by thick lines) and massless external one. Arrows represent momentum direction. The external momentum $p$ obeys the relation $p^2=0$.}
\end{figure}
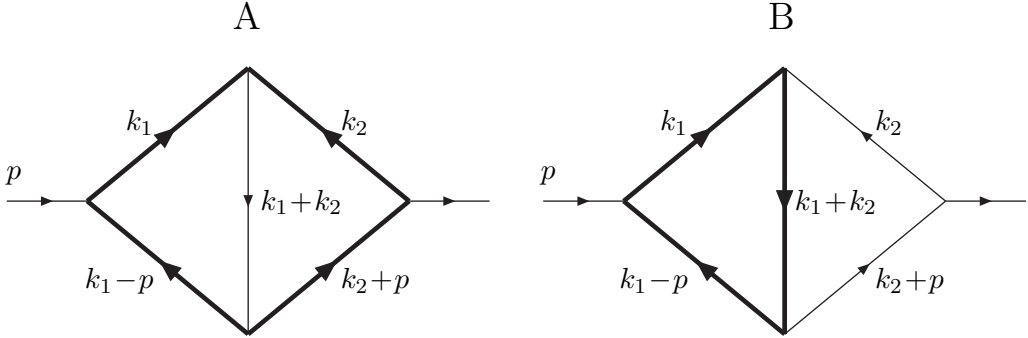

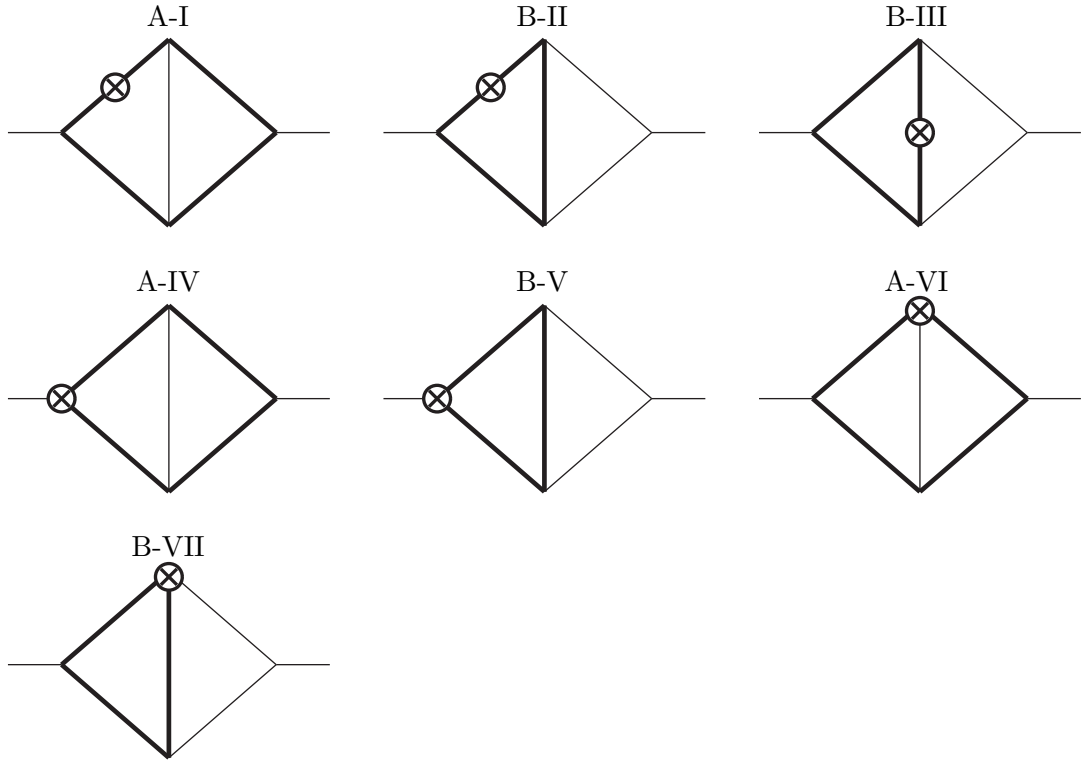
\begin{figure}[tp]
\begin{center}
\begin{picture}(400,290)(0,20)

\Line(000,235)(020,235)
{\SetWidth{1.75} \Line(020,235)(060,270)}
{\SetWidth{1.75} \Line(020,235)(060,200)}
\Line(060,270)(060,200)
{\SetWidth{1.75} \Line(060,270)(100,235)}
{\SetWidth{1.75} \Line(060,200)(100,235)}
\Line(100,235)(120,235)
{\SetWidth{1.25}
\CCirc(40,252){5}{Black}{White}
\Line(37,249)(43,255)
\Line(37,255)(43,249)
}
\Text(060,280)[c]{A-I}

\Line(140,235)(160,235)
{\SetWidth{1.75} \Line(160,235)(200,270)}
{\SetWidth{1.75} \Line(160,235)(200,200)}
{\SetWidth{1.75} \Line(200,270)(200,200)}
\Line(200,270)(240,235)
\Line(200,200)(240,235)
\Line(240,235)(260,235)
{\SetWidth{1.25}
\CCirc(180,252){5}{Black}{White}
\Line(177,249)(183,255)
\Line(177,255)(183,249)
}
\Text(200,280)[c]{B-II}

\Line(280,235)(300,235)
{\SetWidth{1.75} \Line(300,235)(340,270)}
{\SetWidth{1.75} \Line(300,235)(340,200)}
{\SetWidth{1.75} \Line(340,270)(340,200)}
\Line(340,270)(380,235)
\Line(340,200)(380,235)
\Line(380,235)(400,235)
{\SetWidth{1.25}
\CCirc(340,235){5}{Black}{White}
\Line(337,232)(343,238)
\Line(337,238)(343,232)
}
\Text(340,280)[c]{B-III}

\Line(000,135)(020,135)
{\SetWidth{1.75} \Line(020,135)(060,170)}
{\SetWidth{1.75} \Line(020,135)(060,100)}
\Line(060,170)(060,100)
{\SetWidth{1.75} \Line(060,170)(100,135)}
{\SetWidth{1.75} \Line(060,100)(100,135)}
\Line(100,135)(120,135)
{\SetWidth{1.25}
\CCirc(020,135){5}{Black}{White}
\Line(17,132)(23,138)
\Line(17,138)(23,132)
}
\Text(060,180)[c]{A-IV}

\Line(140,135)(160,135)
{\SetWidth{1.75} \Line(160,135)(200,170)}
{\SetWidth{1.75} \Line(160,135)(200,100)}
{\SetWidth{1.75} \Line(200,170)(200,100)}
\Line(200,170)(240,135)
\Line(200,100)(240,135)
\Line(240,135)(260,135)
{\SetWidth{1.25}
\CCirc(160,135){5}{Black}{White}
\Line(157,132)(163,138)
\Line(157,138)(163,132)
}
\Text(200,180)[c]{B-V}

\Line(280,135)(300,135)
{\SetWidth{1.75} \Line(300,135)(340,170)}
{\SetWidth{1.75} \Line(300,135)(340,100)}
\Line(340,170)(340,100)
{\SetWidth{1.75} \Line(340,170)(380,135)}
{\SetWidth{1.75} \Line(340,100)(380,135)}
\Line(380,135)(400,135)
{\SetWidth{1.25}
\CCirc(340,168){5}{Black}{White}
\Line(337,165)(343,171)
\Line(337,171)(343,165)
}
\Text(340,180)[c]{A-VI}

\Line(000,035)(020,035)
{\SetWidth{1.75} \Line(020,035)(060,070)}
{\SetWidth{1.75} \Line(020,035)(060,000)}
{\SetWidth{1.75} \Line(060,070)(060,000)}
\Line(060,070)(100,035)
\Line(060,000)(100,035)
\Line(100,035)(120,035)
{\SetWidth{1.25}
\CCirc(60,68){5}{Black}{White}
\Line(57,65)(63,71)
\Line(57,71)(63,65)
}
\Text(060,080)[c]{B-VII}

\end{picture} 
\end{center}
\caption{\label{InsertionDiagrams} Comprehensive list of two-triangle loop diagrams with operator insertion stemming from the calculation of
the heavy-quark effects in the renormalized operator matrix elements of DIS at $O(\alpha_s^2)$.}
\end{figure}

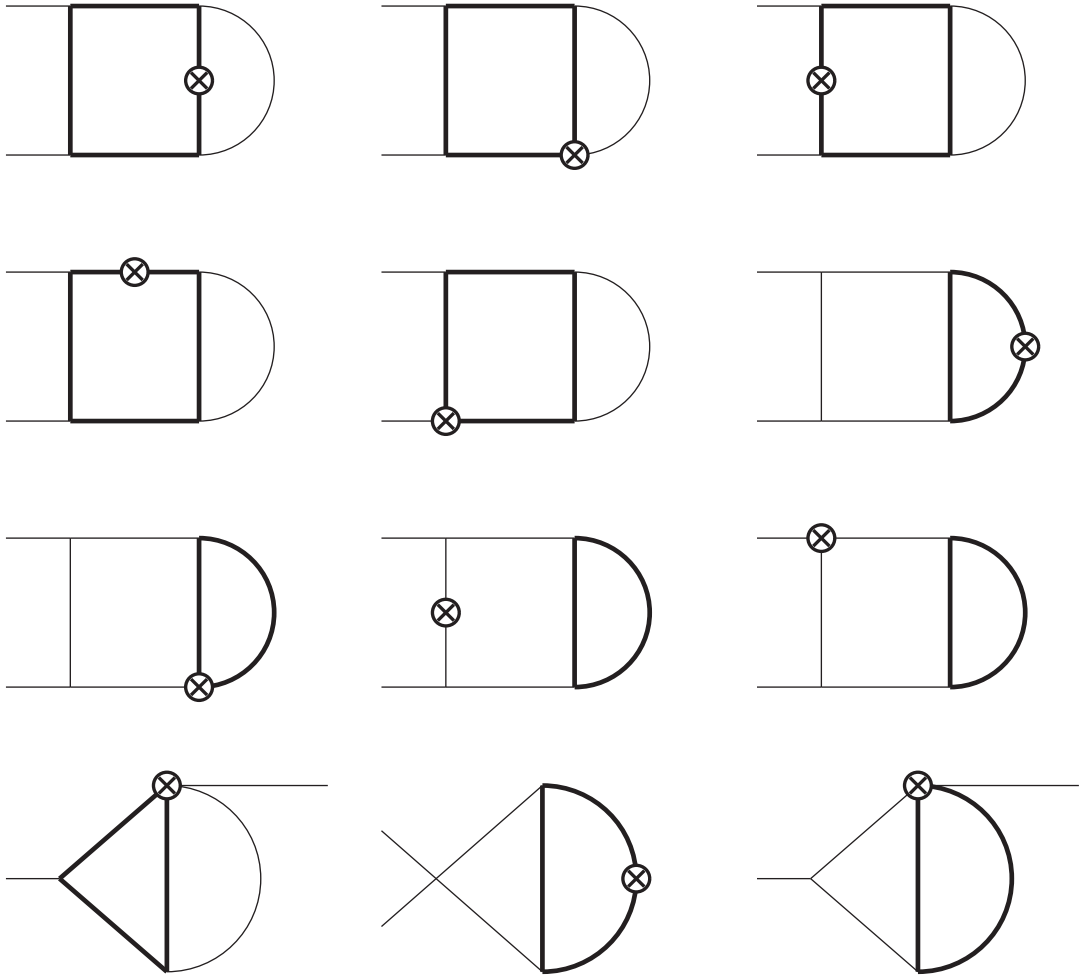
\begin{figure}[tp]
\begin{center}
\begin{picture}(400,370)(0,20)

\Line(000,307)(024,307)
\Line(000,363)(024,363)
{\SetWidth{1.75} 
\Line(024,363)(072,363)
\Line(024,307)(072,307)
\Line(024,307)(024,363)
\Line(072,363)(072,307)
}
\CArc(072,335)(028,-90,90)
{\SetWidth{1.25}
\CCirc(072,335){5}{Black}{White}
\Line(069,332)(075,338)
\Line(069,338)(075,332)
}

\Line(140,307)(164,307)
\Line(140,363)(164,363)
{\SetWidth{1.75} 
\Line(164,363)(212,363)
\Line(164,307)(212,307)
\Line(164,307)(164,363)
\Line(212,363)(212,307)
}
\CArc(212,335)(028,-90,90)
{\SetWidth{1.25}
\CCirc(212,307){5}{Black}{White}
\Line(209,304)(215,310)
\Line(209,310)(215,304)
}

\Line(280,307)(304,307)
\Line(280,363)(304,363)
{\SetWidth{1.75} 
\Line(304,363)(352,363)
\Line(304,307)(352,307)
\Line(304,307)(304,363)
\Line(352,363)(352,307)
}
\CArc(352,335)(028,-90,90)
{\SetWidth{1.25}
\CCirc(304,335){5}{Black}{White}
\Line(301,332)(307,338)
\Line(301,338)(307,332)
}


\Line(000,207)(024,207)
\Line(000,263)(024,263)
{\SetWidth{1.75} 
\Line(024,263)(072,263)
\Line(024,207)(072,207)
\Line(024,207)(024,263)
\Line(072,263)(072,207)
}
\CArc(072,235)(028,-90,90)
{\SetWidth{1.25}
\CCirc(048,263){5}{Black}{White}
\Line(045,260)(051,266)
\Line(045,266)(051,260)
}

\Line(140,207)(164,207)
\Line(140,263)(164,263)
{\SetWidth{1.75} 
\Line(164,263)(212,263)
\Line(164,207)(212,207)
\Line(164,207)(164,263)
\Line(212,263)(212,207)
}
\CArc(212,235)(028,-90,90)
{\SetWidth{1.25}
\CCirc(164,207){5}{Black}{White}
\Line(161,204)(167,210)
\Line(161,210)(167,204)
}

\Line(280,207)(304,207)
\Line(280,263)(304,263)
\Line(304,263)(352,263)
\Line(304,207)(352,207)
\Line(304,207)(304,263)
{\SetWidth{1.75} 
\Line(352,263)(352,207)
\CArc(352,235)(028,-90,90)
}
{\SetWidth{1.25}
\CCirc(380,235){5}{Black}{White}
\Line(377,232)(383,238)
\Line(377,238)(383,232)
}


\Line(000,107)(024,107)
\Line(000,163)(024,163)
\Line(024,163)(072,163)
\Line(024,107)(072,107)
\Line(024,107)(024,163)
{\SetWidth{1.75} 
\Line(072,163)(072,107)
\CArc(072,135)(028,-90,90)
}
{\SetWidth{1.25}
\CCirc(072,107){5}{Black}{White}
\Line(069,104)(075,110)
\Line(069,110)(075,104)
}

\Line(140,107)(164,107)
\Line(140,163)(164,163)
\Line(164,163)(212,163)
\Line(164,107)(212,107)
\Line(164,107)(164,163)
{\SetWidth{1.75} 
\Line(212,163)(212,107)
\CArc(212,135)(028,-90,90)
}
{\SetWidth{1.25}
\CCirc(164,135){5}{Black}{White}
\Line(161,132)(167,138)
\Line(161,138)(167,132)
}

\Line(280,107)(304,107)
\Line(280,163)(304,163)
\Line(304,163)(352,163)
\Line(304,107)(352,107)
\Line(304,107)(304,163)
{\SetWidth{1.75} 
\Line(352,163)(352,107)
\CArc(352,135)(028,-90,90)
}
{\SetWidth{1.25}
\CCirc(304,163){5}{Black}{White}
\Line(301,160)(307,166)
\Line(301,166)(307,160)
}


\Line(000,035)(020,035)
{\SetWidth{1.75} 
\Line(020,035)(060,070)
\Line(020,035)(060,000)
\Line(060,070)(060,000)
}
\Line(060,070)(120,070)
\CArc(060,035)(35,-90,90)
{\SetWidth{1.25}
\CCirc(060,070){5}{Black}{White}
\Line(057,067)(063,073)
\Line(057,073)(063,067)
}

\Line(140,017)(200,070)
\Line(140,053)(200,000)
{\SetWidth{1.75} \Line(200,070)(200,000)}
{\SetWidth{1.75} \CArc(200,035)(35,-90,90)}
{\SetWidth{1.25}
\CCirc(235,035){5}{Black}{White}
\Line(232,032)(238,038)
\Line(232,038)(238,032)
}

\Line(280,035)(300,035)
\Line(300,035)(340,070)
\Line(300,035)(340,000)
{\SetWidth{1.75} \Line(340,070)(340,000)}
\Line(340,070)(400,070)
{\SetWidth{1.75} \CArc(340,035)(35,-90,90)}
{\SetWidth{1.25}
\CCirc(340,070){5}{Black}{White}
\Line(337,067)(343,073)
\Line(337,073)(343,067)
}

\end{picture} 
\end{center}
\caption{\label{OtherDiagramsWithInsertion} Remaining two-loop diagrams necessary for the calculation of the heavy-quark effects in the renormalized matrix elements of DIS at $O(\alpha_s^2)$. Because of the presence of a bubble loop, these calculations are generally simpler than those presented in figure \ref{InsertionDiagrams}.}
\end{figure}

In practice we will have to calculate two-loop integrals with massive internal particles and operator insertions with Feynman rules given in figure \ref{InsertionFeynmanRules} (in order to compare with \cite{Bierenbaum:2006mq} we adopt Mellin-space notation). 
We will start the calculation with two pedagogical examples that do not include operator insertions, as shown in figure \ref{DiagramsWithoutInsertion}. These diagrams are similar to self-energy corrections to a massless particle, however the external momentum obeys the identity $p^2=0$. We then proceed to the calculation of diagrams with operator insertions, involved in the $O(\alpha_s^2)$ corrections to the operator matrix elements. These are separated in two groups, those that do not involve a bubble loop as a building block and those that do, as shown in figures \ref{InsertionDiagrams} and \ref{OtherDiagramsWithInsertion}. We will focus our attention on the first group since these lead to the most complicated integrals. Regarding the second group, the presence of a bubble loop generally leads to simpler calculations and so we will not be performing them explicitly, although we did verify that the method may be applied in all cases. These diagrams represent only the integration part of the calculation, and so we do not specify which particles appear in each diagram, but only whether they are massive or not.

\subsection{Diagrams without operator insertions}

In this subsection we will calculate the integrations originating from diagrams shown in figure \ref{DiagramsWithoutInsertion}. These calculations do not include operator insertions and are similar to self-energy corrections to a massless particle, however with external momentum $p^2=0$. This subsection is included as a stepping stone in understanding the integrations with operator insertions to be performed in subsection \ref{Diagrams with operator insertions}.

For clarity, we list here the steps we will be following:\\
\parbox{14.5cm}{
\begin{itemize}
\item perform first momentum integration and express result in terms of an artificial propagator in second integration;
\vspace{-2mm} \item perform second momentum integration;
\vspace{-2mm} \item expand parameter integrand using Taylor series and perform integration;
\vspace{-2mm} \item expand gamma functions in powers of $\eps$ and apply Z-Sum reduction algorithms;
\vspace{-2mm} \item obtain result in terms of polylogarithms.
\end{itemize}
}

\subsubsection{Diagram A}

We start with an explicit calculation of the integration for diagram A (figure \ref{DiagramsWithoutInsertion}). Since we are interested only in the integration, we ignore all prefactors, which potentially include coupling constants, color factors, and invariants. The initial expression is given by:
\ba
&& \int \frac{d^D k_2}{\(2 \pi\)^D} \frac{1}{\(k_2^2-m^2\)^{\nu_4} \(\(k_2+p\)^2-m^2\)^{\nu_5} } \\
&\times& \int \frac{d^D k_1}{\(2 \pi\)^D} \frac{1}{\(k_1^2-m^2\)^{\nu_1} \(\(k_1+k_2\)^2\)^{\nu_2} \(\(k_1-p\)^2-m^2\)^{\nu_3} } \. \nn
\ea
We perform the $k_1$ integration and re-express the result in terms of an artificial propagator:
\ba
&& \frac{i (-1)^{\nu_{123}}}{\(4 \pi\)^\frac{D}{2}} \frac{\Gamma\!\(\nu_{123}-\frac{D}{2}\)}{\Gamma\!\(\nu_1\) \Gamma\!\(\nu_2\) \Gamma\!\(\nu_3\)} 
\int_0^1 \frac{dx\ \delta_{123}\ x_1^{\nu_1-1}\ x_2^{\nu_2-1}\ x_3^{\nu_3-1} }{\(-x_2 (1-x_2) \)^{\nu_{123}-\frac{D}{2}} }\nl
&\times& \int \frac{d^D k_2}{\(2 \pi\)^D} \frac{1}{\(k_2^2-m^2\)^{\nu_4} \(\(k_2+p\)^2-m^2\)^{\nu_5} \(\(k_2+\frac{x_3\ p}{1-x_2}\)^2-\frac{m^2}{x_2}\)^{\nu_{123}-\frac{D}{2}} }\,
\ea
where $\delta_{123}=\delta(1-x_1-x_2-x_3)$ and $dx$ represents the integration on all $x_\ell$ ($\ell=1,\ 2,\ 3$). At this point the $k_2$ integration may be performed to obtain:
\ba
&& \frac{(-1)^{\nu_{12345}+1}}{\(4 \pi\)^D \(m^2\)^{\nu_{12345}-D}} \frac{\Gamma\!\(\nu_{12345}-D\)}{\Gamma\!\(\nu_1\) \Gamma\!\(\nu_2\) \Gamma\!\(\nu_3\) \Gamma\!\(\nu_4\) \Gamma\!\(\nu_5\)}\\
&\times& \int_0^1 \frac{dx\ \delta_{123}\ \delta_{456}\ x_1^{\nu_1-1}\ x_2^{\nu_{245}-\frac{D}{2}-1}\ (1\!-\!x_2)^{-\nu_{123}+\frac{D}{2}}\ x_3^{\nu_3-1}\
x_4^{\nu_4-1}\ x_5^{\nu_5-1}\ x_6^{\nu_{123}-\frac{D}{2}-1} }{\( x_2 \(1\!-\!x_6\) + x_6 \)^{\nu_{12345}-D} } \. \nn
\ea
The integration delta functions are removed by applying a change of parameter variables:
\ba
&& \frac{(-1)^{\nu_{12345}+1}}{\(4 \pi\)^D \(m^2\)^{\nu_{12345}-D}} \frac{\Gamma\!\(\nu_{12345}-D\)}{\Gamma\!\(\nu_1\) \Gamma\!\(\nu_2\) \Gamma\!\(\nu_3\) \Gamma\!\(\nu_4\) \Gamma\!\(\nu_5\)}
\int_0^1 du\ u_1^{-\nu_2+\frac{D}{2}-1}\ (1\!-\!u_1)^{\nu_{245}-\frac{D}{2}-1}\nl
&\times& \frac{ 
u_2^{\nu_1-1}\ (1\!-\!u_2)^{\nu_3-1}\ 
u_3^{-\nu_4-1}\ (1\!-\!u_3)^{\nu_5-1}\ 
u_4^{-\nu_{45}-1}\ (1\!-\!u_4)^{\nu_{123}-\frac{D}{2}-1}\  
}{
\( 1-u_1\ u_4 \)^{\nu_{12345}-D} 
} \.
\ea
The denominator may be expanded after the inclusion of a regulator $\alpha$, as in $(1-u_1 u_4) \to (1-\alpha u_1 u_4)$, leading to:
\ba
\label{C5-SumA}
&& \frac{(-1)^{\nu_{12345}+1}}{\(4 \pi\)^D \(m^2\)^{\nu_{12345}-D}}
\frac{\Gamma\!\(\nu_{245}-\frac{D}{2}\) \Gamma\!\(\nu_{123}-\frac{D}{2}\)}{\Gamma\!\(\nu_2\) \Gamma\!\(\nu_{13}\) \Gamma\!\(\nu_{45}\)} \\
&\times& \sum_{i=0}^\infty \alpha^i\ \frac{\Gamma\!\(i+\nu_{12345}-D\)}{\Gamma\!\(i+1\)} \frac{\Gamma\!\(i-\nu_2+\frac{D}{2}\)}{\Gamma\!\(i+\nu_{12345}-\frac{D}{2}\)} \.\nn
\ea
Equation (\ref{C5-SumA}) may be expressed in terms of Z-Sums and ultimately polylogarithms even for generic values of $\nu_\ell$ ($\ell=1,\ldots,5$) and $D$ by using identities \ref{PartialFractioningofGammaPair} and \ref{PartialFractioningofGammaPair2}. However for simplicity we will finish this calculation using the specific values of $\nu_\ell=1$ ($\ell=1,\ldots,5$) and $D=4-2\eps$. The expression in eq.(\ref{C5-SumA}) becomes:
\ba
\label{SumDiagramA}
&& \frac{1}{\(4 \pi\)^4 m^2} \(\frac{4 \pi}{m^2}\)^{2 \eps} \Gamma\!\(1+\eps\)^2\  
\sum_{i=0}^\infty \alpha^i\ \frac{\Gamma\!\(i+1+2\eps\)}{\Gamma\!\(i+1\)} \frac{\Gamma\!\(i+1-\eps\)}{\Gamma\!\(i+3+\eps\)}\.
\ea
It is interesting to notice that this expression is simpler than the one obtained when using Mellin-Barnes splitting \cite{Bierenbaum:2006mq}, as it involves a single infinite summation, with both leading to the same result.

Since eq. (\ref{SumDiagramA}) is finite in $\eps$, we apply the limit $\eps \to 0$, leading to:
\ba
&& \frac{1}{\(4 \pi\)^4 m^2} \sum_{i=0}^\infty \alpha^i\  \frac{\Gamma\!\(i+1\)}{\Gamma\!\(i+3\)} \nl
&=& \frac{1}{\(4 \pi\)^4 m^2} \sum_{i=0}^\infty \alpha^i\ \(\frac{1}{i+1}-\frac{1}{i+2}\)\\
&=& \frac{1}{\(4 \pi\)^4 m^2} \(\frac{1}{\alpha} - \log\(1-\alpha\) \(\frac{1}{\alpha}-\frac{1}{\alpha^2} \) \) \.\nn
\ea
The final step is to remove the regulator $\alpha$. In the limit $\alpha \to 1$ the last terms vanish and we obtain:
\be
\frac{1}{\(4 \pi\)^4 m^2}\.
\ee
The simplicity of the result is due to the fact that this is a two point function with massless on-shell external particle.

\subsubsection{Diagram B}

The calculation for diagram B (figure \ref{DiagramsWithoutInsertion}) follows the same procedure and so we omit the intermediate equations. The initial expression is given by:
\ba
&& \int \frac{d^D k_2}{\(2 \pi\)^D} \frac{1}{\(k_2^2-m^2\)^{\nu_4} \(\(k_2+p\)^2-m^2\)^{\nu_5} } \\
&\times& \int \frac{d^D k_1}{\(2 \pi\)^D} \frac{1}{\(k_1^2\)^{\nu_1} \(\(k_1+k_2\)^2-m^2\)^{\nu_2} \(\(k_1-p\)^2\)^{\nu_3} } \.\nn
\ea
We proceed by performing the momentum integrations, expanding the denominator after inclusion of a regulator $\alpha$, and finally performing the parameter integrations. We obtain:
\ba
\label{Beq2}
&& \frac{(-1)^{\nu_{12345}+1}}{\(4 \pi\)^D \(m^2\)^{\nu_{12345}-D}}
\frac{\Gamma\!\(\nu_{1345}-\frac{D}{2}\) \Gamma\!\(\nu_{123}-\frac{D}{2}\)}{\Gamma\!\(\nu_2\) \Gamma\!\(\nu_{13}\) \Gamma\!\(\nu_{45}\)} \\
&\times& \sum_{i=0}^\infty \alpha^i\ \frac{\Gamma\!\(i+\nu_{12345}-D\)}{\Gamma\!\(i+1\)} \frac{\Gamma\!\(i-\nu_{13}+\frac{D}{2}\)}{\Gamma\!\(i+\nu_{12345}-\frac{D}{2}\)}\.\nn
\ea
After setting $\nu_\ell=1$ ($\ell=1,\ldots,5$) and $D=4-2 \eps$ the expression in eq. (\ref{Beq2}) becomes:
\ba
\frac{\Gamma\!\(2+\eps\) \Gamma\!\(1+\eps\) }{\(4 \pi\)^{4-2\eps} \(m^2\)^{1+2\eps}} 
\sum_{i=0}^\infty \alpha^i\ \frac{\Gamma\!\(i+1+2\eps\)}{\Gamma\!\(i+1\)} \frac{\Gamma\!\(i-\eps\)}{\Gamma\!\(i+3+\eps\)}\.
\ea
Unlike diagram A, this expression is not finite so we need to expand the gamma functions in powers of $\eps$. This can be done using the method discussed in section \ref{Gamma Function Expansion and Z-Sums}, after which we obtain:
\be
\frac{\Gamma\!\(2+\eps\) \Gamma\!\(1+\eps\) }{\(4 \pi\)^{4-2\eps} \(m^2\)^{1+2\eps}} 
\(\frac{-1}{2\eps}+\frac{3}{2}-\frac{1}{2 \alpha} -\frac{1}{2 \alpha} \(1-\frac{2}{\alpha}+\frac{1}{\alpha^2}\) \log\(1-\alpha\)\)\.
\ee
When we take the limit $\alpha \to 1$ the last term vanishes and the final result becomes:
\be
\frac{\Gamma\!\(2+\eps\) \Gamma\!\(1+\eps\) }{\(4 \pi\)^{4-2\eps} \(m^2\)^{1+2\eps}} 
\(-\frac{1}{2\eps}+1\)\.
\ee

\subsection{Diagrams with operator insertions}
\label{Diagrams with operator insertions}

In this subsection we will reproduce results presented in \cite{Bierenbaum:2006mq} by calculating diagrams involving operator insertions, shown in figure \ref{InsertionDiagrams}, first obtained by \cite{Buza:1995ie}. We perform the calculation explicitly for two examples, and also present results for the remaining ones.

\subsubsection{Momentum integration with insertion operators}

Before we start with specific calculations let us examine how the new factors in the numerator, originating from the modified Feynman rules of figure \ref{InsertionFeynmanRules}, affect a general triangle integration. Consider a prototype integration of the form:
\be
\int \frac{d^D k}{\(2 \pi\)^D} \frac{\(\Delta \cdot k \)^s}{\(k^2-m_1^2\)^{\nu_1}} \frac{1}{\(\(k-a_2\)^2-m_2^2\)^{\nu_2}} \frac{1}{\(\(k-a_3\)^2-m_3^2\)^{\nu_3}}\.
\ee
The initial steps, that is, introduction of Feynman parameters and completion of the square in the denominator, may be applied as before and we obtain:
\be
\frac{\Gamma\!\(\nu_{123}\)}{\Gamma\!\(\nu_{1}\)\Gamma\!\(\nu_{2}\)\Gamma\!\(\nu_{3}\)}
\int_0^1 dx\ \delta_{123}\ x_1^{\nu_1-1}\ x_2^{\nu_2-1}\ x_3^{\nu_3-1}
\int \frac{d^D k'}{\(2 \pi\)^D} \frac{\(\Delta \cdot \(k'+x_2 a_2 +x_3 a_3\) \)^s}{\({k'}^2-\Delta\)^{\nu_{123}}}\.
\ee
Before we can proceed, we need to expand the factor in the numerator using binomials:
\ba
&&\frac{\Gamma\!\(\nu_{123}\)}{\Gamma\!\(\nu_{1}\)\Gamma\!\(\nu_{2}\)\Gamma\!\(\nu_{3}\)} \sum_{j_1=0}^s \binom{s}{j_1} \sum_{j_2=0}^{j_1} \binom{j_1}{j_2} \(\Delta \cdot a_2\)^{j_1-j_2} \(\Delta \cdot a_3\)^{j_2} \\
&\times& \int_0^1 dx\ \delta_{123}\ x_1^{\nu_1-1}\ x_2^{j_1-j_2+\nu_2-1}\ x_3^{j_2+\nu_3-1} 
\int \frac{d^D k'}{\(2 \pi\)^D} \frac{\(\Delta \cdot k' \)^{s-j_1}}{\({k'}^2-\Delta\)^{\nu_{123}}}\nn\.
\ea
Because $\Delta^2=0$, terms involving non-zero powers of $\left(\Delta \cdot k' \right)^{s-j_1}$ inside the integration vanish, and so our expression becomes
\ba
&&\frac{\Gamma\!\(\nu_{123}\)}{\Gamma\!\(\nu_{1}\)\Gamma\!\(\nu_{2}\)\Gamma\!\(\nu_{3}\)} \sum_{j_2=0}^{s} \binom{s}{j_2} \(\Delta \cdot a_2\)^{s-j_2} \(\Delta \cdot a_3\)^{j_2} \\
&\times& \int_0^1 dx\ \delta_{123}\ x_1^{\nu_1-1}\ x_2^{s-j_2+\nu_2-1}\ x_3^{j_2+\nu_3-1} 
\int \frac{d^D k'}{\(2 \pi\)^D} \frac{1}{\({k'}^2-\Delta\)^{\nu_{123}}}\nn\.
\ea
At this point the regular procedure may be resumed. The end result is simply the normal expression for a triangle loop summed over powers of $\Delta\!\cdot\!a$, as in
\be
I_{cc}^*= \sum_{j=0}^{s} \binom{s}{j} \(\Delta \cdot a_2\)^{s-j} \(\Delta \cdot a_3\)^{j} I_{cc}\ \ ,
\ee
where $I_{cc}^*$ and $I_{cc}$ represent the triangle integration with and without insertion operators, respectively.

\subsubsection{Diagram B-II}

The integral for diagram B-II (figure \ref{InsertionDiagrams}) is given by:
\ba
&& \int \frac{d^D k_2}{\(2 \pi\)^D} \frac{\(\Delta \cdot k_2\)^{N-1}}{\(k_2^2-m^2\)^{\nu_4} \(\(k_2+p\)^2-m^2\)^{\nu_5} } \\
&\times& \int \frac{d^D k_1}{\(2 \pi\)^D} \frac{1}{\(k_1^2\)^{\nu_1} \(\(k_1+k_2\)^2-m^2\)^{\nu_2} \(\(k_1-p\)^2\)^{\nu_3} } \nn\.
\ea
The $k_1$ integration does not involve operator insertions and so it is performed as usual leading to:
\ba
&& \frac{i (-1)^{\nu_{123}}}{\(4 \pi\)^\frac{D}{2}} \frac{\Gamma\!\(\nu_{123}-\frac{D}{2}\)}{\Gamma\!\(\nu_1\) \Gamma\!\(\nu_2\) \Gamma\!\(\nu_3\)} 
\int_0^1 \frac{dx\ \delta_{123}\ x_1^{\nu_1-1}\ x_2^{\nu_2-1}\ x_3^{\nu_3-1} }{\(-x_2 (1-x_2) \)^{\nu_{123}-\frac{D}{2}} }\\
&\times& \int \frac{d^D k_2}{\(2 \pi\)^D} \frac{\(\Delta \cdot k_2\)^{N-1}}{\(k_2^2-m^2\)^{\nu_4} \(\(k_2+p\)^2-m^2\)^{\nu_5} \(\(k_2+\frac{x_3\ p}{1-x_2}\)^2-\frac{m^2}{x_2}\)^{\nu_{123}-\frac{D}{2}} }\nn\.
\ea
Following the steps discussed in the previous subsection, we perform the $k_2$ integration involving the insertion operator and obtain:
\ba
&&\frac{\(-1\)^{\nu_{12345}+1} \(-\Delta \cdot p\)^{N-1}}{\(4 \pi\)^D \(m^2\)^{\nu_{12345}-D}}
\frac{\Gamma\!\(\nu_{12345}-D\)}{\Gamma\!\(\nu_1\) \Gamma\!\(\nu_2\) \Gamma\!\(\nu_3\) \Gamma\!\(\nu_4\) \Gamma\!\(\nu_5\)} \sum_{j=0}^{N-1} \binom{N-1}{j}\\
&\times&\!\!\!\! \int \frac{dx\ \delta_{123}\ \delta_{456}\
x_1^{\nu_1-1} x_2^{-\nu_{13}+\frac{D}{2}-1} \(1\!-\!x_2\)^{-j+\nu_{45}-\frac{D}{2}} x_3^{j+\nu_{3}-1}
x_4^{\nu_{4}-1} x_5^{N-j+\nu_{5}-2} x_6^{j+\nu_{123}-\frac{D}{2}-1}
}{\(\(1-x_6\)\(1-x_2\)+x_6\)^{\nu_{12345-D}}}\nn.
\ea
Applying a change of integration variables and introducing a regulator $\alpha$ we get:
\ba
&&\frac{\(-1\)^{\nu_{12345}+1} \(-\Delta \cdot p\)^{N-1}}{\(4 \pi\)^D \(m^2\)^{\nu_{12345}-D}}
\frac{\Gamma\!\(\nu_{12345}-D\)}{\Gamma\!\(\nu_1\) \Gamma\!\(\nu_2\) \Gamma\!\(\nu_3\) \Gamma\!\(\nu_4\) \Gamma\!\(\nu_5\)} \sum_{j=0}^{N-1} \binom{N-1}{j}\nl
&\times& \int du\ u_1^{-\nu_{13}+\frac{D}{2}-1}\ \(1-u_1\)^{\nu_{1345}-\frac{D}{2}-1}\ u_2^{\nu_{1}-1}\ \(1-u_2\)^{j+\nu_{3}-1}\\
&\times& \frac{u_3^{\nu_4-1}\ \(1-u_3\)^{N-j+\nu_{5}-2}\ u_4^{N-j+\nu_{45}-2}\ \(1-u_4\)^{j+\nu_{123}-\frac{D}{2}-1}
}{\(1-\alpha\ u_1\ u_4\)^{\nu_{12345-D}}}\nn\.
\ea
After expanding the denominator and performing the parameter integrations we obtain:
\ba
&&\frac{\(-1\)^{\nu_{12345}+1} \(-\Delta \cdot p\)^{N-1}}{\(4 \pi\)^D \(m^2\)^{\nu_{12345}-D}}
\frac{\Gamma\!\(\nu_{1345}-\frac{D}{2}\)}{\Gamma\!\(\nu_2\) \Gamma\!\(\nu_3\) \Gamma\!\(\nu_5\)} \nl
&\times& \sum_{j=0}^{N-1} \binom{N-1}{j} \frac{\Gamma\!\(j+\nu_3\)}{\Gamma\!\(j+\nu_{13}\)} \frac{\Gamma\!\(N-j+\nu_5-1\)}{\Gamma\!\(N-j+\nu_{45}-1\)} \\
&\times& \sum_{i=0}^\infty \alpha^i \frac{\Gamma\!\(i+\nu_{12345}-D\)}{\Gamma\!\(i+1\)} \frac{\Gamma\!\(i-\nu_{13}+\frac{D}{2}\)}{\Gamma\!\(i+\nu_{45}\)}
\frac{\Gamma\!\(i+N-j+\nu_{45}-1\) \Gamma\!\(j+\nu_{123}-\frac{D}{2}\)}{\Gamma\!\(i+N+\nu_{12345}-\frac{D}{2}-1\)}\nn\.
\ea
Using eqs. (\ref{PartialFractioningofGammaPair}) and (\ref{PartialFractioningofGammaPair2}) it is possible to systematically express the above equation in terms of Z-Sums and ultimately polylogarithms for generic values of $N$, $\nu_\ell$, and $D$. For simplicity, we will calculate it for $\nu_\ell=1$ ($\ell=1,\ldots,5$) and $D=4-2\eps$. Apart from the prefactor $\frac{\left(-\Delta\!\cdot\!p\right)^{N-1}}{\left(4 \pi\right)^{4-2\eps} \left(m^2\right)^{1+2\eps}}$ we get:
\ba
\label{ResultB-II}
&& \Gamma\!\(2+\eps\) 
\sum_{j=0}^{N-1} \binom{N-1}{j} \frac{\Gamma\!\(j+1\) \Gamma\!\(j+1+\eps\) \Gamma\!\(N-j\)}{\Gamma\!\(j+2\) \Gamma\!\(N-j+1\)}\nl
&\times& \sum_{i=0}^\infty \alpha^i \frac{\Gamma\!\(i-\eps\) \Gamma\!\(i+1-2\eps\) \Gamma\!\(i-j+N+1\)}{\Gamma\!\(i+1\) \Gamma\!\(i+2\) \Gamma\!\(i+N+2+\eps\)}\.
\ea
At this point the Z-Sum reduction algorithms may be applied, similarly to what was done for the cases without operator insertions. We obtain an expression in terms of polylogarithms, apply the limit $\alpha \to 1$ to remove the regulator, and obtain the final results, which, for this particular case, no longer depend on polylogarithms. This calculation has been performed for $N$ equal to 2 through 5 in order to reproduce previous results obtained by other methods \cite{Bierenbaum:2006mq}, and are presented in table \ref{InsertionOperatorResults}.

\subsubsection{Diagram A-VI}

Diagram A-VI involves the insertion of a vertex operator, and so it is a bit more complicated than the previous example. We start the calculation with:
\ba
&& \sum_{j_1=0}^{N-2} \int \frac{d^D k_2}{\(2 \pi\)^D} \frac{\(-\Delta \cdot k_2\)^{N-j_1-1}}{\(k_2^2-m^2\)^{\nu_4} \(\(k_2+p\)^2-m^2\)^{\nu_5} } \\
&\times& \int \frac{d^D k_1}{\(2 \pi\)^D} \frac{\(\Delta \cdot k_1\)^{j_1}}{\(k_1^2-m^2\)^{\nu_1} \(\(k_1+k_2\)^2\)^{\nu_2} \(\(k_1-p\)^2-m^2\)^{\nu_3} } \nn\.
\ea
In this case both integrations include a factor in the numerator. After the $k_1$ integration we obtain:
\ba
&&  \frac{i (-1)^{\nu_{123}}}{\(4 \pi\)^\frac{D}{2}} \frac{\Gamma\!\(\nu_{123}-\frac{D}{2}\)}{\Gamma\!\(\nu_1\) \Gamma\!\(\nu_2\) \Gamma\!\(\nu_3\)} 
\sum_{j_1=0}^{N-2} \sum_{j_2=0}^{j_1} \binom{j_1}{j_2} (-1)^{N-j_1+j_2-2} \(\Delta \cdot p\)^{j_1-j_2}\nl
&\times& \int_0^1 \frac{dx\ \delta_{123}\ x_1^{\nu_1-1}\ x_2^{\nu_2-1}\ x_3^{\nu_3-1} }{\(-x_2 \(1\!-\!x_2\)\)^{\nu_{123}-\frac{D}{2}}}\\
&\times& \int \frac{d^D k_2}{\(2 \pi\)^D} \frac{\(\Delta \cdot k_2\)^{N-j_1+j_2-2}}{\(k_2^2-m^2\)^{\nu_4} 
\(\(k_2+p\)^2-m^2\)^{\nu_5}  \(\(k_2+\frac{x_3 p}{1-x_2}\)^2-\frac{m^2}{x_2}\)^{\nu_{123}-\frac{D}{2}}}\nn\.
\ea
We proceed with the $k_2$ integration and change parameter variables:
\ba
&& \frac{(-1)^{\nu_{12345}+1}}{\(4 \pi\)^D} \frac{\(\Delta \cdot p\)^{N-2}}{\(m^2\)^{\nu_{12345}-D}}
\frac{\Gamma\!\(\nu_{12345}-D\)}{\Gamma\!\(\nu_1\) \Gamma\!\(\nu_2\) \Gamma\!\(\nu_3\) \Gamma\!\(\nu_4\)  \Gamma\!\(\nu_5\)}\nl
&\times& \sum_{j_1=0}^{N-2} \sum_{j_2=0}^{j_1} \binom{j_1}{j_2} \sum_{j_3=0}^{N\!-\!j_1\!+\!j_2\!-\!2} \binom{N\!-\!j_1\!+\!j_2\!-\!2}{j_3} \\
&\times& \int_0^1 du\
u_1^{j_1-j_2-\nu_2+\frac{D}{2}-1}\ \(1\!-\!u_1\)^{j_2+\nu_{245}-\frac{D}{2}-1}\
u_2^{\nu_1-1}\ \(1\!-\!u_2\)^{j_1-j_2+j_3+\nu_3-1}\nl
&\times& \frac{
u_3^{\nu_4-1}\ \(1\!-\!u_3\)^{N-j_1+j_2-j_3+\nu_5-3}\
u_4^{N-j_1+j_2-j_3+\nu_{45}-3}\ \(1\!-\!u_4\)^{j_3+\nu_{123}-\frac{D}{2}-1}  
}{\(1-\alpha\ u_1\ u_4\)^{\nu_{12345}-D}}\nn\.
\ea
After denominator expansion and parameter integration we get:
\ba
&& \frac{(-1)^{\nu_{12345}+1}}{\(4 \pi\)^D} \frac{\(\Delta \cdot p\)^{N-2}}{\(m^2\)^{\nu_{12345}-D}}
\frac{1}{\Gamma\!\(\nu_2\) \Gamma\!\(\nu_3\) \Gamma\!\(\nu_5\)} \sum_{j_1=0}^{N-2} \sum_{j_2=0}^{j_1} \binom{j_1}{j_2} \\
&\times&  \sum_{j_3=0}^{N-j_1+j_2-2} \binom{N\!-\!j_1\!+\!j_2\!-\!2}{j_3} 
\frac{\Gamma\!\(j_1\!-\!j_2\!+\!j_3\!+\!\nu_3\)}{\Gamma\!\(j_1\!-\!j_2\!+\!j_3\!+\!\nu_{13}\)} \frac{\Gamma\!\(N\!-\!j_1\!+\!j_2\!-\!j_3\!+\!\nu_5\!-\!2\)}{\Gamma\!\(N\!-\!j_1\!+\!j_2\!-\!j_3\!+\!\nu_{45}\!-\!2\)}\nl
&\times& \sum_{i=0}^\infty \alpha^i \frac{\Gamma\!\(i\!+\!\nu_{12345}\!-\!D\)}{\Gamma\!\(i\!+\!1\)}
\frac{\Gamma\!\(i\!+\!j_1\!-\!j_2\!-\!\nu_2\!+\!\frac{D}{2}\) \Gamma\!\(j_2\!+\!\nu_{245}\!-\!\frac{D}{2}\)}{\Gamma\!\(i\!+\!j_1\!+\!\nu_{45}\)}\nl
&\times& \frac{\Gamma\!\(i\!+\!N\!-\!j_1\!+\!j_2\!-\!j_3\!+\!\nu_{45}\!-\!2\) \Gamma\!\(j_3\!+\!\nu_{123}\!-\!\frac{D}{2}\)}{\Gamma\!\(i\!+\!N\!-\!j_1\!+\!j_2\!+\!\nu_{12345}\!-\!2\)}\nn\.
\ea
For the purpose of this example we set $\nu_\ell=1$ ($\ell=1,\ldots,5$) and $D=4$. Apart from a prefactor we get:
\ba
\label{ResultA-VI}
&&\sum_{j_1=0}^{N-2} \sum_{j_2=0}^{j_1} \binom{j_1}{j_2} \sum_{j_3=0}^{N-j_1+j_2-2} \binom{n-j_1+j_2-2}{j_3}
\frac{\Gamma\!\(j_2+1\) \Gamma\!\(j_3+1\) \Gamma\!\(j_1-j_2+j_3+1\) }{\Gamma\!\(j_1-j_2+j_3+2\) }\nl
&\times& \frac{\Gamma\!\(N-j_1+j_2-j_3-1\)}{\Gamma\!\(N-j_1+j_2-j_3\)} 
\sum_{i=0}^\infty \alpha^i \frac{\Gamma\!\(i+j_1-j_2+1\) \Gamma\!\(i+N-j_1+j_2-j_3\)}{\Gamma\!\(i+j_1+2\) \Gamma\!\(i+N-j_1+j_2+1\)}\.
\ea
At this point the Z-Sum reduction algorithms may be applied, similarly to what was done previously. We obtain an expression in terms of polylogarithms, apply the limit $\alpha \to 1$ to remove the regulator, and obtain the final results, which, for this particular case, no longer depend on polylogarithms. Results for $N$ equal to 2 through 5 are presented in table \ref{InsertionOperatorResults}.

\subsubsection{Remaining diagrams}

Having explicitly discussed two calculations as examples, we end this section by presenting results for the remaining operator-insertion diagrams required in the calculation of the process. Calculations for all diagrams in figure \ref{InsertionDiagrams} have been reproduced including the first four Mellin moments and are presented in table \ref{InsertionOperatorResults}. In order to match the previous results \cite{Bierenbaum:2006mq}, we did not include a constant prefactor, similarly to what was done to obtain eqs. (\ref{ResultB-II}) (diagram B-II) and (\ref{ResultA-VI}) (diagrams A-VI).

{\renewcommand{\arraystretch}{2.2}
\renewcommand{\tabcolsep}{0.1cm}
\begin{table}[tp]
\begin{center}
{\small
\begin{tabular}{|c|c|c|c|c|}
\hline
N  	& $2$										& $3$												& $4$										& $5$								\\ \hline

A-Ia 	& $\df{1}{2}$									& $\df{67}{216}$										& $\df{31}{144}$								& $\df{2161}{13500}$						\\ \hline

A-Ib	& $-\df{13}{144}$								& $-\df{19}{432}$										& $-\df{17}{675}$								& $-\df{431}{27000}$						\\ \hline

B-II 	& $-\df{1}{4\eps}+\df{1}{4}+\df{\gamma_E}{2}$					& $-\df{11}{72\eps}+\df{23}{144}+\df{11\gamma_E}{36}$						& $-\df{5}{48 \eps}+\df{11}{96}+\df{5\gamma_E}{24}$				& $-\df{137}{1800\eps}+\df{949}{10800}+\df{137 \gamma_E}{900}$	\\ \hline

A-III	& $0$										& $-\df{1}{24\eps}+\df{1}{48}+\df{\gamma_E}{12}$						& $0$										& $-\df{1}{90\eps}+\df{1}{270}+\df{\gamma_E}{45}$		\\ \hline

A-IV 	& $1$										& $0$												& $\df{31}{72}$									& $0$								\\ \hline

B-V  	& $-\df{1}{2\eps}+\df{1}{2}+\gamma_E$						& $0$												& $-\df{5}{24\eps}+\df{11}{48}+\df{5\gamma_E}{12}$  				& $0$								\\ \hline

A-VI 	& $1$										& $1$												& $\df{65}{72}$									& $\df{29}{36}$							\\ \hline

B-VII	& $-\df{1}{2\eps}+\df{1}{2}+\gamma_E$						& $-\df{1}{4\eps}+\df{1}{4}+\df{\gamma_E}{2}$							& $-\df{5}{24\eps}+\df{29}{144}+\df{5\gamma_E}{12}$				& $-\df{5}{36\eps}+\df{7}{48}+\df{5\gamma_E}{18}$		\\ \hline
\end{tabular}
}
\caption{
Results for the first four Mellin momenta for the diagrams shown in figure \ref{InsertionDiagrams}.\label{InsertionOperatorResults}}
\end{center}
\end{table}
}

While we will not be presenting results for diagrams involving bubbles as building blocks (figure \ref{OtherDiagramsWithInsertion}), because of their simplicity, we note that the method can be successfully applied in all cases.

\section{Conclusions}
\label{Conclusions}
We have developed a study of different methods to systematically perform expansions of multi-loop calculations leading to concatenated sums, with the intent of applying the Z-Sum algorithms in order to obtain final results in terms of multiple polylogarithms. These methods involved the use of Taylor expansions or inverse Mellin-Barnes transformations, applied either at the level of momentum integration or parameter integration. While most of the focus in the literature in recent years has been on the application of approach C (see table \ref{TableABC}), based on inverse Mellin-Barnes transformations, to specific calculations, we decided to cover and develop a way to systematically use Taylor series expansions to obtain a general parameter integration. This allowed us to clearly understand the structure of the summations obtained, and thus evaluate how often the method can be successfully applied, while at the same time making progress towards a generalization of the method by identifying missing algorithms.

We proceeded by surveying the applicability of the Taylor expansion method to one-loop diagrams in QFT and found that not all calculations could be performed using the Z-Sum machinery as it is, specially if several external legs are off-shell. Since some of these diagrams are used as building blocks for multi-loop calculations this demonstrates the limitations of the procedure, with similar difficulties to be expected for box diagrams. While this is a concern if the ultimate goal is to develop an automatic computational package capable of performing higher-loop calculations, it does not necessarily put this approach in a worse position than the alternatives, given that such a general survey has not yet been done for the other methods and so their limitations are still unknown.

Since one would like to explore the possibility of expressing results in terms of multiple polylogarithms, we see two different approaches in proceeding with this line of research. One would be to try to extend the Z-Sum machinery in order to obtain the necessary steps for an entirely systematic reduction using approach A, finding solutions to eqs. (\ref{MA1}-\ref{MA4}) and their generalizations. The other option would be to forgo the use of Taylor series and, similarly to what was done in this work, develop a systematic procedure for performing a Mellin-Barnes expansion of a general loop integration using approaches B or C. This would allow us to understand the general form of the concatenated sums obtained from such methods, finding out how often they can be successfully applied, and which new algorithms are necessary for a general solution.

Finally, having discovered in which situations approach A can be successfully used, in section \ref{Application} we conclude by applying it to a problem of physical interest, namely the calculation of heavy-flavor effects in the structure functions of deep inelastic scattering, which involves the calculation of two-loop diagrams with massive internal particles and operator insertions. This is an important result since for the first time multi-loop calculations involving massive internal particles have been performed using the Z-Sum reduction algorithms and Taylor expansions. By doing so, we were able to reproduce results obtained using inverse Mellin-Barnes transformations, as presented in \cite{Bierenbaum:2006mq}.

This calculation shows that there are in fact cases of multi-loop diagrams involving massive particles where the method may be successfully applied. We believe this work sets a useful foundation for further investigation into applications of Z-Sum algorithms for calculations of loop integrals.

\nocite{*}

\acknowledgments This work is supported in part by the U.S. Department of Energy under grant DE-FG02-97IR41022, and by the National Science Foundation under Grant No. PHY05-51164. P. Rottmann would like to thank Isabella Bierenbaum and Stefan Weinzierl for helpful comments during the preparation of \cite{Rottmann:2011Diss}, on which this paper in based.

\newpage
\appendix
\section{Notation and conventions}
\label{Notation and Conventions}

\paragraph{Boolean step function}
Whenever we use the step function $\theta$ it will have a boolean argument as in:
\begin{equation}
  \theta(x) = \left\{
    \begin{array}{ll}
      1		& \text{if } x = \text{true}\\
      0		& \text{if } x = \text{false}\\
          \end{array} \right.\.
\end{equation}

\paragraph{Floor and ceiling function}
The floor function $\lfloor x \rfloor$ evaluates as the largest integer smaller or equal to $x$. Similarly, the ceiling function $\lceil x \rceil$ equals the smallest integer larger or equal to $x$. Namely:
\ba
\lfloor x \rfloor &=& n ,\4 n \in \mathbb{I},\4 0 \le x-n < 1\,\\
\lceil x \rceil &=& n ,\4 n \in \mathbb{I},\4 0 \le n-x < 1\.\nn
\ea

\section{Summation splitting and reordering}
\label{SumsReordering}

When dealing with concatenated sums, it is useful and sometimes even necessary to shift the summation variable and its limits, or reorder sums. An example of such a requirement is when one needs to compare two different functions by expressing them in the same basis of monomials. In this appendix we list some useful identities that we have repeatedly used in obtaining results presented in sections \ref{Triangle Loop Integrals} and \ref{Application}.

\subsection{Basic identities with a single sum}
\label{Basic Identities with a Single Sum}
We start with very basic ones involving one sum only:
\be
\sum_{i=0}^n f(i) = \sum_{i=0}^n f(n-i) \,
\ee

\be
\label{SumAddingTerms1}
\sum_{j=0}^{n+i} f(j) = \sum_{j=0}^i f(j)+ \sum_{j=0}^n f(j+i) - f(i) \,
\ee

\be
\label{SumAddingTerms2}
\sum_{j=0}^{n-i} f(j) = \sum_{j=0}^n f(j)+ \sum_{j=0}^i f(j+n-i) - f(n-i) \,
\ee
where $f(i)$ represents any function of $i$, and n is an integer.

In eq. (\ref{SumAddingTerms2}) we added and subtracted terms originally
outside the range of the original summation, which might lead to badly defined terms (for example involving
$\Gamma(-1)$). These will eventually cancel out but a regularizer should be used so that intermediate steps are
well defined.

It is possible to express a sum in terms of other summations with multiplied upper limit. A trivial example comes from splitting into odd and even sums:
\be
\sum_{i=0}^{n}\ f(i) = \sum_{i=0}^{\lfloor \frac{n}{2} \rfloor}\ f(2i)\ +\ \sum_{i=0}^{\lfloor \frac{n-1}{2} \rfloor}\ f(2i+1)\ ,
\ee
with the inverted identity given by:
\be
\label{doubbleupperlimit}
\sum_{i=1}^n f(i) = \frac{1}{2} \(\sum_{i=1}^{2n}\ f(i/2) + \sum_{i=1}^{2n}\ (-1)^i\ f(i/2) \)\ ,
\ee
where odd terms simply cancel out.

These equations can be generalized to:
\be
\label{GeneralLimitMultiply1}
\sum_{i=0}^{n} f(i) = \sum_{p=0}^{q-1} \sum_{i=0}^{\lfloor \frac{n-p}{q} \rfloor} f(q i + p) \,
\ee
and
\be
\label{GeneralLimitMultiply2}
\sum_{i=1}^n f(i)=\frac{1}{q} \sum_{p=0}^{q-1} \sum_{i=1}^{q n} (r_q^p)^i f(i/q)\ ,
\ee
where $p$ and $q$ are integers and the coefficient $r_q^p$ is called root of unity \cite{Weinzierl:2004bn}, and is defined by:
\be
r_q^p = \exp\(\frac{2 \pi i p}{q}\) \,
\ee
with the properties:
\be
\(r_q^p\)^{j+q} = \(r_q^p\)^{j} \,
\ee
and
\be
  \sum_{p=0}^{q-1} \(r_q^p\)^m= \left\{
    \begin{array}{rl}
      q & \text{if } m \text{ mod } q = 0 \,\\
      0 & \text{if } m \text{ mod } q \neq 0 \.
    \end{array} \right.
\ee

\subsection{Splitting identities}
The multiplication of Z-Sums is a result of sum splitting:
\be
\label{ConcSumSplitting}
\sum_{i=1}^n \sum_{j=1}^n f(i,j) = \sum_{i=1}^n \sum_{j=1}^{i-1} f(i,j) + \sum_{j=1}^n \sum_{i=1}^{j-1} f(i,j) 
+ \sum_{i=1}^n f(i,i) \ .
\ee
A minor limit modification (useful for sums originating from denominator Taylor series) gives:
\be
\sum_{i=0}^n \sum_{j=0}^n f(i,j) = \sum_{i=0}^n \sum_{j=0}^{i} f(i,j) + \sum_{j=0}^n \sum_{i=0}^{j} f(i,j) 
- \sum_{i=0}^n f(i,i) \ .
\ee

Concatenated sums involving floor function in the upper limit may be split into odd and even parts according to:
\be
\sum_{i=0}^a \sum_{j=0}^{\lfloor \frac{i+b}{2} \rfloor +c} f(i,j) =
\sum_{i=0}^{\lfloor \frac{a}{2} \rfloor} \2 \sum_{j=0}^{i+\lfloor \frac{b}{2} \rfloor +c} f(2i,j)+
\sum_{i=0}^{\lfloor \frac{a-1}{2} \rfloor} \2 \sum_{j=0}^{i+\lfloor \frac{b+1}{2} \rfloor +c} f(2i+1,j)\,
\ee
where $a$, $b$, and $c$ are integer numbers.

\subsection{Shifting identities}
Other identities obtained from sum reordering are:
\ba
\label{ShiftArgFin1}
&& \sum_{i=0}^n \sum_{j=0}^i f(i,j) = \sum_{i=0}^n \sum_{j=0}^{n-i} f(i+j,j) = \sum_{i=0}^n \sum_{j=0}^{n-i} f(i+j,i)\,\\
\label{ShiftArgFin2}
&& \sum_{i=0}^n \sum_{j=0}^{n-i} f(i,j) = \sum_{i=0}^n \sum_{j=0}^{n-i} f(j,i) = \sum_{i=0}^n \sum_{j=0}^{i} f(i-j,j)\,\\
\label{ShiftArgInf1}
&& \sum_{i=0}^\infty \sum_{j=0}^i f(i,j) = \sum_{i=0}^\infty \sum_{j=0}^\infty f(i+j,j) = \sum_{i=0}^\infty \sum_{j=0}^\infty f(i+j,i)\,\\
\label{ShiftArgInf2}
&& \sum_{i=0}^\infty \sum_{j=0}^\infty f(i,j) = \sum_{i=0}^\infty \sum_{j=0}^\infty f(j,i) = \sum_{i=0}^\infty \sum_{j=0}^{i} f(i-j,j)\ .
\ea
Equation (\ref{ShiftArgFin2}) is simply eq. (\ref{ShiftArgFin1}) rewritten with $f(x,y)\to f(x-y,y)$, while the last two are obtained from the first ones after taking the limit $n \to \infty$.

Similarly to eqs. (\ref{ShiftArgInf1}) and (\ref{ShiftArgInf2}), further shifting leads to:
\be
\sum_{i=0}^\infty \sum_{j=0}^i f(i,j) = \sum_{i=0}^\infty \sum_{j=0}^{\lfloor\frac{i}{2}\rfloor} f(i-j,j)
= \sum_{i=0}^\infty \sum_{j=0}^{\lfloor\frac{i}{m}\rfloor} f(i-(m-1)j,j)\,
\ee
and
\be
\sum_{i=0}^\infty \sum_{j=0}^i f(i,j) = \sum_{i=0}^\infty \sum_{j=0}^{2 i} f\!\left(i+\lfloor\frac{j}{2}\rfloor,j\right)
= \sum_{i=0}^\infty \sum_{j=0}^{m\ i} f\!\left(i-(j-1)-\lfloor\frac{(j-1)}{m}\rfloor,j\right)\ .
\ee
All infinite summations being reordered must be absolutely convergent for these identities to be valid.

\subsection{Reordering of concatenated sums}
Now we present the algorithm for reordering two sums where the inner sum upper and/or lower limits depend linearly on the outer sum's
variable. Since this identity does not change the arguments within the sum we omit $f(i,j)$. Expressions
involving more than two sums can be reordered iteratively, two sums at a time.

The initial sum is given by:
\be
\sum_{i=a}^b \sum_{j=c i+d}^{e i+f} \,
\ee
where $a$, $b$, $c$, $d$, $e$, and $f$ are real numbers. We can express the limits of this summation as:
\ba
\label{ConditionBlock}
a \le & i & \le b \, \\
c i + d \le & j & \le e i +f \, \nonumber
\ea
where $i$ and $j$ are integers. We would like to invert the condition (\ref{ConditionBlock}) and express the summation limits in the form:
\ba
\label{GenericBlock}
f_1(a,b,c,d,e,f) \le & j & \le f_2(a,b,c,d,e,f) \, \\
f_3(j,a,b,c,d,e,f) \le & i & \le f_4(j,a,b,c,d,e,f) \, \nonumber
\ea
which will be rewritten as
\be
\sum_{j=f_1(a,b,c,d,e,f)}^{f_2(a,b,c,d,e,f)}\ \ \sum_{i=f_3(j,a,b,c,d,e,f)}^{f_4(j,a,b,c,d,e,f)}\ .
\ee

In case we obtain a block of the form (\ref{GenericBlock}) involving $<$ or $>$ instead of $\le$ or $\ge$
we make the substitutions $\alpha < i \to \lfloor\alpha+1\rfloor \le i$ and $i < \alpha \to i \le \lceil\alpha-1\rceil$,
which is correct whether $\alpha$ is an integer or not. In all other cases when the summation limits are not integers the
ceiling function for the lower limit and the floor function for the upper limit are implied. 

We separate the solution in eight cases depending on the values of $c$ and $e$. All cases will have constant
blocks for all values of $a$, $b$, $c$, $d$, $e$, and $f$ and also conditional blocks with a boolean step function.

\paragraph{Case 1}$e=0 \land c>0$
\be
\sum_{i=a}^b \sum_{j=c i+d}^{f}
= \sum_{j=\lfloor c b +d+1\rfloor}^f \sum_{i=a}^{b}
+ \theta(c b + d \le f) \sum_{j=c a +d}^{c b +d} \sum_{i=a}^{\frac{j-d}{c}}
+ \theta(c b + d > f) \sum_{j=c a +d}^{f} \sum_{i=a}^{\frac{j-d}{c}}\,
\ee

\paragraph{Case 2}$e=0 \land c<0$
\be
\sum_{i=a}^b \sum_{j=c i+d}^{f}
= \sum_{j=c a + d}^f \sum_{i=a}^{b}
+ \theta(c a + d \le f) \sum_{j=c b +d}^{\lceil c a +d-1 \rceil} \sum_{i=\frac{j-d}{c}}^{b}
+ \theta(c a + d > f) \sum_{j=c b +d}^{f} \sum_{i=\frac{j-d}{c}}^{b}\,
\ee

\paragraph{Case 3}$c=0 \land e>0$
\be
\sum_{i=a}^b \sum_{j=d}^{e i+f}
= \sum_{j=d}^{\lceil e a+f-1\rceil} \sum_{i=a}^{b}
+ \theta(e a + f \le d) \sum_{j=d}^{e b+f} \sum_{i=\frac{j-f}{e}}^{b}
+ \theta(e a + f > d) \sum_{j=e a +f}^{e b+f} \sum_{i=\frac{j-f}{e}}^{b}\,
\ee

\paragraph{Case 4}$c=0 \land e<0$
\be
\sum_{i=a}^b \sum_{j=d}^{e i+f}
= \sum_{j=d}^{\lceil e b+f-1\rceil} \sum_{i=a}^{b}
+ \theta(e b + f \le d) \sum_{j=d}^{e a+f} \sum_{i=a}^{\frac{j-f}{e}}
+ \theta(e b + f > d) \sum_{j=e b +f}^{e a+f} \sum_{i=a}^{\frac{j-f}{e}}\,
\ee

\paragraph{Case 5} $c>0 \land e>0$
\ba
\sum_{i=a}^b \sum_{j=c i+d}^{e i+f}
&=& \sum_{j=\lfloor e a+f+1\rfloor}^{c b+d} \sum_{i=\frac{j-f}{e}}^{\frac{j-d}{c}}
+ \sum_{j=\lfloor c b+d+1\rfloor}^{e a+f} \sum_{i=a}^{b}\nonumber \\
&+& \theta(c b+d \le e a+f)
\left(\sum_{j=c a+d}^{c b+d} \sum_{i=a}^{\frac{j-d}{c}}+\sum_{j=\lfloor e a+f+1 \rfloor}^{e b+f} \sum_{i=\frac{j-f}{e}}^b\right) \\
&+& \theta(c b+d > e a+f)
\left(\sum_{j=c a+d}^{e a+f} \sum_{i=a}^{\frac{j-d}{c}}+\sum_{j=\lfloor c b+d+1 \rfloor}^{e b+f} \sum_{i=\frac{j-f}{e}}^b\right)\,\nonumber
\ea

\paragraph{Case 6}$c>0 \land e<0$
\ba
\sum_{i=a}^b \sum_{j=c i+d}^{e i+f}
&=& \sum_{j=e b+f}^{\frac{c f-e d}{c-e}} \sum_{i=a}^{\frac{j-d}{c}}
+ \sum_{j=\lfloor \frac{c f - e d}{c-e} +1 \rfloor}^{c b+d} \sum_{i=a}^{\frac{j-f}{e}}
+ \sum_{j=\lfloor c b+d+1\rfloor}^{\frac{c f-e d}{c-e}} \sum_{i=a}^{b}
+ \sum_{j=\lfloor \frac{c f - e d}{c-e} +1\rfloor}^{\lceil e b+f-1 \rceil} \sum_{i=a}^{b}\nonumber \\
&+& \theta(c b+d \ge e b+f)
\left(\sum_{j=c a+d}^{\lceil e b+f-1 \rceil} \sum_{i=a}^{\frac{j-d}{c}}+\sum_{j=\lfloor c b+d+1 \rfloor}^{e a+f} \sum_{i=a}^{\frac{j-f}{e}}\right) \\
&+& \theta(c b+d < e b+f)
\left(\sum_{j=c a+d}^{c b+d} \sum_{i=a}^{\frac{j-d}{c}}+\sum_{j=e b+f}^{e a+f} \sum_{i=a}^{\frac{j-f}{e}}\right)\,\nonumber
\ea

\paragraph{Case 7}$c<0 \land e>0$
\ba
\sum_{i=a}^b \sum_{j=c i+d}^{e i+f}
&=& \sum_{j=\frac{c f-e d}{c-e}}^{e a+f} \sum_{i=a}^{b}
+ \sum_{j=c a+d}^{\lceil \frac{c f-e d}{c-e}-1 \rceil} \sum_{i=a}^{b}
+ \sum_{j=\frac{c f-e d}{c-e}}^{\lceil c a+d-1 \rceil} \sum_{i=\frac{j-f}{e}}^{b}
+ \sum_{j=\lfloor e a+f+1\rfloor}^{\lceil \frac{c f-e d}{c-e}-1 \rceil} \sum_{i=\frac{j-d}{c}}^{b}\nonumber \\
&+& \theta(c a+d \le e a+f)
\left(\sum_{j=\lfloor e a+f+1 \rfloor}^{e b+f} \sum_{i=\frac{j-f}{e}}^b+\sum_{j=c b+d}^{\lceil c a+d-1 \rceil} \sum_{i=\frac{j-d}{c}}^b\right) \\
&+& \theta(c a+d > e a+f)
\left(\sum_{j=c a+d}^{e b+f} \sum_{i=\frac{j-f}{e}}^b+\sum_{j=c b+d}^{e a+f} \sum_{i=\frac{j-d}{c}}^b\right)\,\nonumber
\ea

\paragraph{Case 8}$c<0 \land e<0$
\ba
\sum_{i=a}^b \sum_{j=c i+d}^{e i+f}
&=& \sum_{j=e b+f}^{\lceil c a+d-1 \rceil} \sum_{i=\frac{j-d}{c}}^{\frac{j-f}{e}}
+ \sum_{j=c a+d}^{\lceil e b+f-1 \rceil} \sum_{i=a}^{b} \nonumber \\
&+& \theta(c a+d \ge e b+f)
\left(\sum_{j=c a+d}^{e a+f} \sum_{i=a}^{\frac{j-f}{e}}+\sum_{j=c b+d}^{\lceil e b+f-1 \rceil} \sum_{i=\frac{j-d}{c}}^b\right)\\
&+& \theta(c a+d < e b+f)
\left(\sum_{j=e b+f}^{e a+f} \sum_{i=a}^{\frac{j-f}{e}}+\sum_{j=c b+d}^{\lceil c a+d-1 \rceil} \sum_{i=\frac{j-d}{c}}^b\right)\.\nonumber
\ea

\bibliographystyle{JHEP}
\bibliography{References}

\providecommand{\href}[2]{#2}\begingroup\raggedright\begin{thebibliography}{10}

\bibitem{Moch:2001zr}
S.~Moch, P.~Uwer, and S.~Weinzierl, {\it {Nested sums, expansion of
  transcendental functions and multi-scale multi-loop integrals}},  {\em J.
  Math. Phys.} {\bf 43} (2002) 3363--3386,
  [\href{http://xxx.lanl.gov/abs/hep-ph/0110083}{{\tt hep-ph/0110083}}].

\bibitem{Weinzierl:2003jx}
S.~Weinzierl, {\it {Algebraic algorithms in perturbative calculations}},
  \href{http://xxx.lanl.gov/abs/hep-th/0305260}{{\tt hep-th/0305260}}.

\bibitem{Moch:2002rq}
S.~Moch, P.~Uwer, and S.~Weinzierl, {\it {QCD two-loop amplitudes for $e^+ e^-
  \to$ 3jets: The fermionic contribution}},  {\em Nucl. Phys. Proc. Suppl.}
  {\bf 116} (2003) 8--12, [\href{http://xxx.lanl.gov/abs/hep-ph/0211156}{{\tt
  hep-ph/0211156}}].

\bibitem{Moch:2002wt}
S.~Moch, P.~Uwer, and S.~Weinzierl, {\it {Scattering amplitudes for $e^+ e^-
  \to$ 3jets at next-to-next- to-leading order QCD}},  {\em Nucl. Phys. Proc.
  Suppl.} {\bf 121} (2003) 37--41,
  [\href{http://xxx.lanl.gov/abs/hep-ph/0210009}{{\tt hep-ph/0210009}}].

\bibitem{Bierenbaum:2003ud}
I.~Bierenbaum and S.~Weinzierl, {\it {The massless two-loop two-point
  function}},  {\em Eur. Phys. J.} {\bf C32} (2003) 67--78,
  [\href{http://xxx.lanl.gov/abs/hep-ph/0308311}{{\tt hep-ph/0308311}}].

\bibitem{Bierenbaum:2005:T}
I.~Bierenbaum, {\em The Massless Two-loop Two-point Function and Zeta Functions
  in Counterterms of Feynman Diagrams}.
\newblock PhD thesis, Universit{\"a}t Mainz, Feb., 2005.

\bibitem{Blumlein:2006mh}
J.~Blumlein, A.~{De Freitas}, W.~L. van Neerven, and S.~Klein, {\it {The
  longitudinal heavy quark structure function F(L)(Q anti-Q) in the region $Q^2
  >> m^2$ at $O(\alpha(s)^3)$}},  {\em Nucl. Phys.} {\bf B755} (2006) 272--285,
  [\href{http://xxx.lanl.gov/abs/hep-ph/0608024}{{\tt hep-ph/0608024}}].

\bibitem{Bierenbaum:2006mq}
I.~Bierenbaum, J.~Blumlein, and S.~Klein, {\it {Evaluating two-loop massive
  operator matrix elements with Mellin-Barnes integrals}},  {\em Nucl. Phys.
  Proc. Suppl.} {\bf 160} (2006) 85--90,
  [\href{http://xxx.lanl.gov/abs/hep-ph/0607300}{{\tt hep-ph/0607300}}].

\bibitem{Bierenbaum:2007dm}
I.~Bierenbaum, J.~Blumlein, and S.~Klein, {\it {Calculation of massive 2-loop
  operator matrix elements with outer gluon lines}},  {\em Phys. Lett.} {\bf
  B648} (2007) 195--200, [\href{http://xxx.lanl.gov/abs/hep-ph/0702265}{{\tt
  hep-ph/0702265}}].

\bibitem{Bierenbaum:2007pn}
I.~Bierenbaum, J.~Blumlein, and S.~Klein, {\it {Two-Loop Massive Operator
  Matrix Elements for Polarized and Unpolarized Deep-Inelastic Scattering}},
  \href{http://xxx.lanl.gov/abs/0706.2738}{{\tt arXiv:0706.2738}}.

\bibitem{Bierenbaum:2007qe}
I.~Bierenbaum, J.~Blumlein, and S.~Klein, {\it {Two-loop massive operator
  matrix elements and unpolarized heavy flavor production at asymptotic values
  $Q^2 >> m^2$}},  {\em Nucl. Phys.} {\bf B780} (2007) 40--75,
  [\href{http://xxx.lanl.gov/abs/hep-ph/0703285}{{\tt hep-ph/0703285}}].

\bibitem{Bierenbaum:2008yu}
I.~Bierenbaum, J.~Blumlein, S.~Klein, and C.~Schneider, {\it {Two--Loop Massive
  Operator Matrix Elements for Unpolarized Heavy Flavor Production to
  $O(\epsilon)$}},  {\em Nucl. Phys.} {\bf B803} (2008) 1--41,
  [\href{http://xxx.lanl.gov/abs/0803.0273}{{\tt arXiv:0803.0273}}].

\bibitem{Bierenbaum:2009zt}
I.~Bierenbaum, J.~Blumlein, and S.~Klein, {\it {The Gluonic Operator Matrix
  Elements at $O(\alpha_s^2)$ for DIS Heavy Flavor Production}},  {\em Phys.
  Lett.} {\bf B672} (2009) 401--406,
  [\href{http://xxx.lanl.gov/abs/0901.0669}{{\tt arXiv:0901.0669}}].

\bibitem{Bierenbaum:2010jp}
I.~Bierenbaum, J.~Blumlein, and S.~Klein, {\it {Logarithmic $O(\alpha_s^3)$
  contributions to the DIS Heavy Flavor Wilson Coefficients at $Q^2 \gg m^2$}},
   {\em PoS} {\bf DIS2010} (2010) 148,
  [\href{http://xxx.lanl.gov/abs/1008.0792}{{\tt arXiv:1008.0792}}].

\bibitem{Bejdakic:2008cr}
E.~Bejdakic, {\it {Multiloop Bubbles for hot QCD}},  {\em Nucl. Phys.} {\bf
  A820} (2009) 263c--266c, [\href{http://xxx.lanl.gov/abs/0810.3097}{{\tt
  arXiv:0810.3097}}].

\bibitem{Heinrich:2009be}
G.~Heinrich, T.~Huber, D.~A. Kosower, and V.~A. Smirnov, {\it {Nine-Propagator
  Master Integrals for Massless Three-Loop Form Factors}},  {\em Phys. Lett.}
  {\bf B678} (2009) 359--366, [\href{http://xxx.lanl.gov/abs/0902.3512}{{\tt
  arXiv:0902.3512}}].

\bibitem{Weinzierl:2009ms}
S.~Weinzierl, {\it {Event shapes and jet rates in electron-positron
  annihilation at NNLO}},  {\em JHEP} {\bf 06} (2009) 041,
  [\href{http://xxx.lanl.gov/abs/0904.1077}{{\tt arXiv:0904.1077}}].

\bibitem{Weinzierl:2009nz}
S.~Weinzierl, {\it {The infrared structure of $e^+ e^- \to 3\ jets$ at NNLO
  reloaded}},  {\em JHEP} {\bf 07} (2009) 009,
  [\href{http://xxx.lanl.gov/abs/0904.1145}{{\tt arXiv:0904.1145}}].

\bibitem{Bolzoni:2009ye}
P.~Bolzoni, S.-O. Moch, G.~Somogyi, and Z.~Trocsanyi, {\it {Analytic
  integration of real-virtual counterterms in NNLO jet cross sections II}},
  {\em JHEP} {\bf 08} (2009) 079,
  [\href{http://xxx.lanl.gov/abs/0905.4390}{{\tt arXiv:0905.4390}}].

\bibitem{Weinzierl:2010cw}
S.~Weinzierl, {\it {Jet algorithms in electron-positron annihilation:
  Perturbative higher order predictions}},  {\em Eur. Phys. J.} {\bf C71}
  (2011) 1565, [\href{http://xxx.lanl.gov/abs/1011.6247}{{\tt
  arXiv:1011.6247}}].

\bibitem{Huber:2010fz}
T.~Huber, {\it {Master integrals for massless three-loop form factors}},
  \href{http://xxx.lanl.gov/abs/1001.3132}{{\tt arXiv:1001.3132}}.

\bibitem{Ablinger:2010ha}
J.~Ablinger {\em et.~al.}, {\it {Heavy Flavor DIS Wilson coefficients in the
  asymptotic regime}},  {\em Nucl. Phys. Proc. Suppl.} {\bf 205-206} (2010)
  242--249, [\href{http://xxx.lanl.gov/abs/1007.0375}{{\tt arXiv:1007.0375}}].

\bibitem{DelDuca:2010zp}
V.~{Del Duca}, C.~Duhr, and V.~A. Smirnov, {\it {A Two-Loop Octagon Wilson Loop
  in N = 4 SYM}},  {\em JHEP} {\bf 09} (2010) 015,
  [\href{http://xxx.lanl.gov/abs/1006.4127}{{\tt arXiv:1006.4127}}].

\bibitem{DelDuca:2010zg}
V.~{Del Duca}, C.~Duhr, and V.~A. Smirnov, {\it {The Two-Loop Hexagon Wilson
  Loop in N = 4 SYM}},  {\em JHEP} {\bf 05} (2010) 084,
  [\href{http://xxx.lanl.gov/abs/1003.1702}{{\tt arXiv:1003.1702}}].

\bibitem{Smirnov:2004}
V.~A. Smirnov, {\em Evaluating Feynman Integrals}.
\newblock Springer, 2004.

\bibitem{Czakon:2005rk}
M.~Czakon, {\it {Automatized analytic continuation of Mellin-Barnes
  integrals}},  {\em Comput. Phys. Commun.} {\bf 175} (2006) 559--571,
  [\href{http://xxx.lanl.gov/abs/hep-ph/0511200}{{\tt hep-ph/0511200}}].

\bibitem{Weinzierl:2004bn}
S.~Weinzierl, {\it {Expansion around half-integer values, binomial sums and
  inverse binomial sums}},  {\em J. Math. Phys.} {\bf 45} (2004) 2656--2673,
  [\href{http://xxx.lanl.gov/abs/hep-ph/0402131}{{\tt hep-ph/0402131}}].

\bibitem{Rottmann:2011Diss}
P.~Rottmann, {\em Z-Sum Approach to Loop Integrals}.
\newblock PhD thesis, Florida State University, 2011.

\bibitem{Euler:1775:M}
L.~Euler, {\it Meditationes circa singulare serierum genus},  {\em Novi Comm.
  Acad. Sci. Petropol.} {\bf 20} (1775) 140--186.

\bibitem{Goncharov:1998:M}
A.~Goncharov, {\it Multiple polylogarithms, cyclotomy and modular complexes},
  {\em Mathematical Research Letters} {\bf 5} (1998), no.~3 497--516.

\bibitem{Lewin:1981:P}
L.~Lewin, {\em Polylogarithms and Associated Functions}.
\newblock Elsevier Science Ltd, 1981.

\bibitem{Nielsen:1909:D}
N.~Nielsen, {\it Der {E}ulersche {D}ilogarithmus und seine
  {V}erallgemeinerungen},  {\em Nova Acta Leopoldina (Halle)} (1909), no.~90
  123.

\bibitem{Remiddi:1999ew}
E.~Remiddi and J.~A.~M. Vermaseren, {\it {Harmonic polylogarithms}},  {\em Int.
  J. Mod. Phys.} {\bf A15} (2000) 725--754,
  [\href{http://xxx.lanl.gov/abs/hep-ph/9905237}{{\tt hep-ph/9905237}}].

\bibitem{vanOldenborgh:1990wj}
G.~J. {van Oldenborgh} and J.~A.~M. Vermaseren, {\it {The formula manipulation
  program Form}}, . Prepared for International Workshop on Software
  Engineering, Artificial Intelligence and Expert Systems for High-energy and
  Nuclear Physics, Lyon, France, 19-24 Mar 1990.

\bibitem{Vermaseren:2000nd}
J.~A.~M. Vermaseren, {\it {New features of FORM}},
  \href{http://xxx.lanl.gov/abs/math-ph/0010025}{{\tt math-ph/0010025}}.

\bibitem{Moch:2005uc}
S.~Moch and P.~Uwer, {\it {XSummer: Transcendental functions and symbolic
  summation in Form}},  {\em Comput. Phys. Commun.} {\bf 174} (2006) 759--770,
  [\href{http://xxx.lanl.gov/abs/math-ph/0508008}{{\tt math-ph/0508008}}].

\bibitem{Buza:1995ie}
M.~Buza, Y.~Matiounine, J.~Smith, R.~Migneron, and W.~L. van Neerven, {\it
  {Heavy quark coefficient functions at asymptotic values $Q~2 \gg m~2$}},
  {\em Nucl. Phys.} {\bf B472} (1996) 611--658,
  [\href{http://xxx.lanl.gov/abs/hep-ph/9601302}{{\tt hep-ph/9601302}}].

\bibitem{2010CoPhC.181..582B}
J.~{Bl{\"u}mlein}, D.~J. {Broadhurst}, and J.~A.~M. {Vermaseren}, {\it {The
  Multiple Zeta Value data mine}},  {\em Computer Physics Communications} {\bf
  181} (Mar., 2010) 582--625, [\href{http://xxx.lanl.gov/abs/0907.2557}{{\tt
  arXiv:0907.2557}}].

\bibitem{Heinrich:2007at}
G.~Heinrich, T.~Huber, and D.~Maitre, {\it {Master Integrals for Fermionic
  Contributions to Massless Three-Loop Form Factors}},  {\em Phys. Lett.} {\bf
  B662} (2008) 344--352, [\href{http://xxx.lanl.gov/abs/0711.3590}{{\tt
  arXiv:0711.3590}}].

\bibitem{Ablinger:2010ty}
J.~Ablinger, J.~Blumlein, S.~Klein, C.~Schneider, and F.~Wissbrock, {\it {The
  $O(\alpha_s^3)$ Massive Operator Matrix Elements of $O(n_f)$ for the
  Structure Function $F_2(x,Q^2)$ and Transversity}},  {\em Nucl. Phys.} {\bf
  B844} (2011) 26--54, [\href{http://xxx.lanl.gov/abs/1008.3347}{{\tt
  arXiv:1008.3347}}].

\bibitem{2011PhLB..698..443K}
D.~{Kreimer} and K.~{Yeats}, {\it {Tensor structure from scalar Feynman
  matroids}},  {\em Physics Letters B} {\bf 698} (Apr., 2011) 443--450,
  [\href{http://xxx.lanl.gov/abs/1010.5804}{{\tt arXiv:1010.5804}}].

\bibitem{Broadhurst:1998rz}
D.~J. Broadhurst, {\it {Massive 3-loop Feynman diagrams reducible to SC*
  primitives of algebras of the sixth root of unity}},  {\em Eur. Phys. J.}
  {\bf C8} (1999) 311--333, [\href{http://xxx.lanl.gov/abs/hep-th/9803091}{{\tt
  hep-th/9803091}}].

\bibitem{Gehrmann:2001jv}
T.~Gehrmann and E.~Remiddi, {\it {Numerical evaluation of two-dimensional
  harmonic polylogarithms}},  {\em Comput. Phys. Commun.} {\bf 144} (2002)
  200--223, [\href{http://xxx.lanl.gov/abs/hep-ph/0111255}{{\tt
  hep-ph/0111255}}].

\bibitem{Gehrmann:2001pz}
T.~Gehrmann and E.~Remiddi, {\it {Numerical evaluation of harmonic
  polylogarithms}},  {\em Comput. Phys. Commun.} {\bf 141} (2001) 296--312,
  [\href{http://xxx.lanl.gov/abs/hep-ph/0107173}{{\tt hep-ph/0107173}}].

\bibitem{Bonciani:2003hc}
R.~Bonciani, P.~Mastrolia, and E.~Remiddi, {\it {Master integrals for the
  2-loop QCD virtual corrections to the forward-backward asymmetry}},  {\em
  Nucl. Phys.} {\bf B690} (2004) 138--176,
  [\href{http://xxx.lanl.gov/abs/hep-ph/0311145}{{\tt hep-ph/0311145}}].

\bibitem{Diakonidis:2008ij}
{\mbox{Th. Diakonidis, J. Fleischer, J. Gluza, K. Kajda, T. Riemann, J.B.
  Tausk}}, {\it {A complete reduction of one-loop tensor 5- and 6-point
  integrals}},  {\em Phys. Rev.} {\bf D80} (2009) 036003,
  [\href{http://xxx.lanl.gov/abs/0812.2134}{{\tt arXiv:0812.2134}}].

\bibitem{Aglietti:2007as}
U.~Aglietti, R.~Bonciani, L.~Grassi, and E.~Remiddi, {\it {The Two Loop Crossed
  Ladder Vertex Diagram with Two Massive Exchanges}},  {\em Nucl. Phys.} {\bf
  B789} (2008) 45--83, [\href{http://xxx.lanl.gov/abs/0705.2616}{{\tt
  arXiv:0705.2616}}].

\bibitem{Aglietti:2004tq}
U.~Aglietti and R.~Bonciani, {\it {Master integrals with 2 and 3 massive
  propagators for the 2-loop electroweak form factor: Planar case}},  {\em
  Nucl. Phys.} {\bf B698} (2004) 277--318,
  [\href{http://xxx.lanl.gov/abs/hep-ph/0401193}{{\tt hep-ph/0401193}}].

\bibitem{Aglietti:2003yc}
U.~Aglietti and R.~Bonciani, {\it {Master integrals with one massive propagator
  for the two- loop electroweak form factor}},  {\em Nucl. Phys.} {\bf B668}
  (2003) 3--76, [\href{http://xxx.lanl.gov/abs/hep-ph/0304028}{{\tt
  hep-ph/0304028}}].

\bibitem{Bonciani:2010ms}
R.~Bonciani, G.~Degrassi, and A.~Vicini, {\it {On the Generalized Harmonic
  Polylogarithms of One Complex Variable}},  {\em Comput. Phys. Commun.} {\bf
  182} (2011) 1253--1264, [\href{http://xxx.lanl.gov/abs/1007.1891}{{\tt
  arXiv:1007.1891}}].

\bibitem{Weinzierl:2006qs}
S.~Weinzierl, {\it {The art of computing loop integrals}},
  \href{http://xxx.lanl.gov/abs/hep-ph/0604068}{{\tt hep-ph/0604068}}.

\bibitem{Zagier:1992:V}
D.~Zagier, {\it Values of zeta functions and their applications},  in {\em
  First European Congress of Mathematics}, vol.~2, pp.~497--512, 1992.

\bibitem{Denner:1993:T}
A.~Denner, {\it Techniques for the calculation of electroweak radiative
  corrections at the one-loop level and results for w-physics at lep 200},
  {\em Fortschritte der Physik/Progress of Physics} {\bf 41} (1993), no.~4
  307--420.

\bibitem{Oleari:1997az}
C.~Oleari, {\it {Next-to-leading-order corrections to the production of
  heavy-flavour jets in e+ e- collisions}},
  \href{http://xxx.lanl.gov/abs/hep-ph/9802431}{{\tt hep-ph/9802431}}.

\bibitem{Dittmaier:2003ej}
S.~Dittmaier, .~M. Kramer, and M.~Spira, {\it {Higgs radiation off bottom
  quarks at the Tevatron and the LHC}},  {\em Phys. Rev.} {\bf D70} (2004)
  074010, [\href{http://xxx.lanl.gov/abs/hep-ph/0309204}{{\tt
  hep-ph/0309204}}].

\bibitem{Bogner:2010kv}
C.~Bogner and S.~Weinzierl, {\it {Feynman graph polynomials}},
  \href{http://xxx.lanl.gov/abs/1002.3458}{{\tt arXiv:1002.3458}}.

\bibitem{Ellis:2007qk}
R.~K. Ellis and G.~Zanderighi, {\it {Scalar one-loop integrals for QCD}},  {\em
  JHEP} {\bf 02} (2008) 002, [\href{http://xxx.lanl.gov/abs/0712.1851}{{\tt
  arXiv:0712.1851}}].

\bibitem{Weinzierl:2003ub}
S.~Weinzierl, {\it {Hopf algebra structures in particle physics}},  {\em Eur.
  Phys. J.} {\bf C33} (2004) s871--s875,
  [\href{http://xxx.lanl.gov/abs/hep-th/0310124}{{\tt hep-th/0310124}}].

\bibitem{Weinzierl:2007cx}
S.~Weinzierl, {\it {Feynman integrals and multiple polylogarithms}},
  \href{http://xxx.lanl.gov/abs/0705.0900}{{\tt arXiv:0705.0900}}.

\bibitem{Fleischer:1997bw}
J.~Fleischer, A.~V. Kotikov, and O.~L. Veretin, {\it {The differential equation
  method: Calculation of vertex- type diagrams with one non-zero mass}},  {\em
  Phys. Lett.} {\bf B417} (1998) 163--172,
  [\href{http://xxx.lanl.gov/abs/hep-ph/9707492}{{\tt hep-ph/9707492}}].

\bibitem{Fleischer:1998nb}
J.~Fleischer, A.~V. Kotikov, and O.~L. Veretin, {\it {Analytic two-loop results
  for selfenergy- and vertex-type diagrams with one non-zero mass}},  {\em
  Nucl. Phys.} {\bf B547} (1999) 343--374,
  [\href{http://xxx.lanl.gov/abs/hep-ph/9808242}{{\tt hep-ph/9808242}}].

\bibitem{Davydychev:2003mv}
A.~I. Davydychev and M.~Y. Kalmykov, {\it {Massive Feynman diagrams and inverse
  binomial sums}},  {\em Nucl. Phys.} {\bf B699} (2004) 3--64,
  [\href{http://xxx.lanl.gov/abs/hep-th/0303162}{{\tt hep-th/0303162}}].

\bibitem{Jegerlehner:2002em}
F.~Jegerlehner, M.~Y. Kalmykov, and O.~Veretin, {\it {MS-bar vs pole masses of
  gauge bosons. II: Two-loop electroweak fermion corrections}},  {\em Nucl.
  Phys.} {\bf B658} (2003) 49--112,
  [\href{http://xxx.lanl.gov/abs/hep-ph/0212319}{{\tt hep-ph/0212319}}].

\end{thebibliography}\endgroup

\end{document}